\documentclass[prd,twocolumn,aps,nofootinbib,showpacs,floats,letterpaper,floatfix,groupedaddress,eqsecnum]{revtex4}

\usepackage[dvips]{graphicx}
\usepackage{amssymb}
\usepackage{amsmath}

\usepackage{color}

\usepackage{dcolumn,epsfig}

\def\nn{\nonumber}

\def\be{\begin{equation}}
\def\ee{\end{equation}}
\def\beq{\begin{eqnarray}}
\def\eeq{\end{eqnarray}}

\def\IL{\relax{\rm I\kern-.18em L}}

\def\nn{\nonumber}
\def\f{\frac}

\newcommand{\hxf}{\hat{\boldsymbol{\chi}}_f}

\newcommand{\hLN}{\hat{\mathbf{L}}_{N}}

\newcommand{\hJ}{\hat{\mathbf{J}}}

\newcommand{\vS}{\mathbf{S}}

\newcommand{\xa}{\boldsymbol{\chi}_1}
\newcommand{\xb}{\boldsymbol{\chi}_2}
\newcommand{\xf}{\boldsymbol{\chi}_f}
\newcommand{\xsum}{\boldsymbol{\chi}_s}
\newcommand{\xdiff}{\boldsymbol{\chi}_a}
\newcommand{\LN}{\mathbf{L}_N}
\newcommand{\Sa}{\mathbf{S}_1}
\newcommand{\Sb}{\mathbf{S}_2}
\newcommand{\hSi}{\hat{\mathbf{S}}_i}
\newcommand{\hSa}{\hat{\mathbf{S}}_1}
\newcommand{\hSb}{\hat{\mathbf{S}}_2}
\newcommand{\dSa}{\dot{\mathbf{S}}_1}
\newcommand{\dSb}{\dot{\mathbf{S}}_2}
\newcommand{\dhLN}{\dot{\hat{\mathbf{L}}}_{N}}

\newcommand{\oaSPFa}{\bar{\boldsymbol{\Omega}}_1}
\newcommand{\oaSPFb}{\bar{\boldsymbol{\Omega}}_2}
\newcommand{\DP}{\Delta \phi}

\newcommand{\Szero}{\mathbf{S}_0}

\def\nn{\nonumber}

\begin{document}

\title{Final spins from the merger of precessing binary black holes}

\author{Michael Kesden} \email{kesden@tapir.caltech.edu}
\affiliation{California Institute of Technology, MC 350-17, 1216 E. California
  Blvd., Pasadena, CA 91125}

\author{Ulrich Sperhake} \email{sperhake@tapir.caltech.edu}
\affiliation{California Institute of Technology, MC 350-17, 1216 E. California
  Blvd., Pasadena, CA 91125}

\author{Emanuele Berti} \email{berti@phy.olemiss.edu}
\affiliation{Department of Physics and Astronomy, The University of
  Mississippi, University, MS 38677-1848, USA}
\affiliation{California Institute of Technology, MC 350-17, 1216 E. California
  Blvd., Pasadena, CA 91125}

\date{February 2010}

\begin{abstract}
The inspiral of binary black holes is governed by gravitational
radiation reaction at binary separations $r \lesssim 1000 M$, yet it
is too computationally expensive to begin numerical-relativity
simulations with initial separations $r \gtrsim 10 M$.  Fortunately,
binary evolution between these separations is well described by
post-Newtonian equations of motion.  We examine how this
post-Newtonian evolution affects the distribution of spin orientations
at separations $r \simeq 10 M$ where numerical-relativity simulations
typically begin.  Although isotropic spin distributions at $r
\simeq 1000 M$ remain isotropic at $r \simeq 10 M$, distributions that are
initially partially aligned with the orbital angular momentum can be
significantly distorted during the post-Newtonian inspiral.  Spin
precession tends to align (anti-align) the binary black hole spins
with each other if the spin of the more massive black hole is
initially partially aligned (anti-aligned) with the orbital angular
momentum, thus increasing (decreasing) the average final spin.  Spin
precession is stronger for comparable-mass binaries, and could produce
significant spin alignment before merger for both supermassive and
stellar-mass black hole binaries. We also point out that precession
induces an intrinsic accuracy limitation ($\lesssim 0.03$ in the
dimensionless spin magnitude, $\lesssim 20^\circ$ in the direction) in
predicting the final spin resulting from the merger of widely
separated binaries.
\end{abstract}



\maketitle

\section{Introduction} \label{S:intro}

The existence of black holes is a fundamental prediction of general
relativity.  Isolated individual black holes are stationary solutions to
Einstein's equations, but binary black holes (BBHs) can inspiral and
eventually merge.  BBH mergers offer a unique opportunity to test general
relativity in the strong-field limit, and as such are a primary science target
for current and future gravitational-wave (GW) observatories like LIGO, VIRGO,
LISA, and the Einstein telescope.  BBH mergers are also important for
cosmology, as they can serve as standard candles to help determine the
geometry and hence energy content of the universe
\cite{Schutz:1986gp,Holz:2005df}.  Astrophysical BBHs are found on at least
two very different mass scales.  Compact objects believed to be stellar-mass
black holes have been observed in binary systems with more luminous
companions.  These black holes are the remnants of massive main-sequence
stars, and binary systems with two such stars may ultimately evolve into BBHs.
On larger scales, supermassive black holes (SBHs) with masses $10^6 \lesssim
M/M_\odot \lesssim 10^9$ reside in the centers of most galaxies.  They can be
observed through their dynamical influence on surrounding gas and stars, and
when accreting as active galactic nuclei (AGN).  SBHs will form binaries as
well, following the merger of two galaxies which each host an SBH at their
center.

In order to merge, BBHs must find a way to shed their orbital angular
momentum.  At large separations, binary SBHs will be escorted inwards by
dynamical friction between their host galaxies \cite{Begelman:1980vb}.  The
BBHs become gravitationally bound when the sum of their masses $M \equiv m_1 +
m_2$ exceeds the mass of gas and stars enclosed by their orbit.  The binary
hardens further by scattering stars on ``loss-cone'' orbits that pass within a
critical radius \cite{Frank:1976uy}, though this scattering may stall at
separations $r \simeq 0.01 - 1$ pc unless these orbits are refilled by stellar
diffusion \cite{Milosavljevic:2001vi}.  Unlike stars, gas can cool to form a
circumbinary disk about the BBHs.  A circumbinary disk of mass $M_d$ and
radius $r_d$ will exert a tidal torque
\begin{equation} \label{E:Td}
T_d \sim \frac{q^2 M_d M}{r} \left( \frac{r}{r_d - r} \right)^3
\end{equation}
on the binary in the limit that the BBH mass ratio $q \equiv m_2/m_1
\leq 1$ is small and $|r_d - r| \ll r$
\cite{Lin:1979,Goldreich:1980wa,Chang:2008}.  Throughout this paper we
use relativists' units in which Newton's constant $G$ and the speed of
light $c$ are unity.  At a sufficiently small separation $r_{\rm GW}$,
the magnitude of this tidal torque will fall below that of the
radiation-reaction torque \cite{Peters:1964zz}
\begin{equation} \label{E:TGW}
T_{\rm GW} = \frac{32 \eta^2 M^{9/2}}{5r^{7/2}}~,
\end{equation}
where $\eta \equiv m_1m_2/M^2$ is the symmetric mass ratio.  Once $T_{\rm GW}
> T_d$, the inspiral of the BBH is dominated by radiation reaction.  The
precise value of $r_{\rm GW}$ depends on the properties of the circumbinary
disk, but an order-of-magnitude estimate is given by \cite{Begelman:1980vb}
\begin{subequations} \label{E:rGW}
  \begin{eqnarray} \label{E:rGWcgs}
r_{\rm GW} &=& (5 \times 10^{16}~{\rm cm}) q^{1/4} M_{8}^{3/4}
\left[ \frac{{\rm min}(t_h, t_{\rm gas})}{10^8~{\rm yr}}
\right]^{1/4} \\ \label{E:rGWrel}
&=& (3000 M) \left( \frac{q}{M_8} \right)^{1/4}
\left[ \frac{{\rm min}(t_h, t_{\rm gas})}{10^8~{\rm yr}}
\right]^{1/4}
  \end{eqnarray}
\end{subequations}
where $M_8$ is the mass of the larger black hole in units of $10^8 M_\odot$,
$t_h$ is the dynamical friction timescale for a hard binary, and $t_{\rm gas}$
is the evolution timescale from gaseous tidal torques.

General relativity completely determines the inspiral of BBH systems from
separations less than $r_{\rm GW}$.  These systems are fully specified by 7
parameters: the mass ratio $q$ and the 3 components of each dimensionless spin
$\boldsymbol{\chi}_{1,2} \equiv {\bf S}_{1,2}/m_{1,2}^2$.  To a good
approximation the individual masses and spin magnitudes $\chi_{1,2} \equiv
|\boldsymbol{\chi}_{1,2}|$ remain constant during the inspiral, so only the
precession of the two spin directions needs to be calculated.  At an initial
separation $r_i = 1000 M \sim r_{\rm GW}$, the binary's orbital speed $v/c \ll
1$ and the spin-precession equations can therefore be expanded in this small
post-Newtonian (PN) parameter.  The PN expansion remains valid until the BBHs
reach a final separation $r_f = 10 M$, after which their evolution can only be
described by fully nonlinear numerical relativity (for more precise
assessments of the validity of the PN expansion for spinning precessing
binaries, see e.g.~\cite{Buonanno:2005xu,Campanelli:2008nk}).  Numerical
relativists can simulate BBH mergers from separations $r_{\rm NR} \simeq r_f$
\cite{Pretorius:2005gq,Campanelli:2005dd,Baker:2005vv}, but these simulations
are too computationally expensive to begin when the binaries are much more
widely separated.  The GWs produced in the merger and the mass, spin, and
recoil velocity of the final black hole depend sensitively on the orientation
of the BBH spins at $r_{\rm NR}$, so it is important to determine what BBH
spin orientations are expected at $r_i$ and whether these orientations are
modified by the PN evolution between $r_i$ and $r_f$.

The answer to the first of these questions comes from astrophysics, not
general relativity.  At very large separations, the two black holes are
unaffected by each other and one would therefore expect an isotropic
distribution of spin directions.  However, an isotropic distribution of spins
at $r_f$ would imply that most mergers would result in a gravitational recoil
of $\sim 1000$ km/s for the final black hole
\cite{Gonzalez:2007hi,Campanelli:2007ew,Dotti:2009vz}.  Recoils this large
would eject SBHs from all but the most massive host galaxies
\cite{Merritt:2004xa}, in seeming contradiction to the observed tight
correlations between SBHs and their hosts
\cite{Magorrian:1997hw,Ferrarese:2000se,Tremaine:2002js}.  This problem can be
avoided if Lense-Thirring precession and viscous torques align the spins of
the BBHs with the accretion disk responsible for their inwards migration
\cite{BP,Bogdanovic:2007hp,Berti:2008af}.  The efficiency of this alignment
depends on the properties of the accretion disk, but $N$-body simulations
using smoothed-particle hydrodynamics (SPH) suggest that the residual
misalignment of the BBH spins with their accretion disk at $r_i$ could
typically be $\sim 10^\circ (30^\circ)$ for cold (hot) accretion disks
\cite{Dotti:2009vz}.

The second question, does the distribution of spin directions change
as the BBHs inspiral from $r_i$ to $r_f$, can be answered by evolving
this distribution over this interval using the PN spin-precession
equations.  We will describe these PN equations and our numerical
solutions to them in Sec.~\ref{S:PN}.  The precession of a given spin
configuration in the PN regime can be understood in terms of the
proximity of that configuration to the nearest spin-orbit resonance.
Schnittman \cite{Schnittman:2004vq} identified a set of equilibrium
spin configurations in which both black hole spins and the orbital angular
momentum lie in a plane, along with the total angular momentum ${\bf
J} = {\bf L} + m_{1}^2 \xa + m_{2}^2 \xb$.  In the absence of
radiation reaction, ${\bf J}$ is conserved.  For these equilibrium
configurations, the spins and orbital angular momentum remain coplanar
and precess jointly about ${\bf J}$ with the angles $\theta_{1,2}$
between ${\bf L}$ and $\boldsymbol{\chi}_{1,2}$ remaining fixed.  The
equilibrium configurations can thus be understood as spin-orbit
resonances since the precession frequencies of ${\bf L}$ and
$\boldsymbol{\chi}_{1,2}$ about ${\bf J}$ are all the same.  Once
radiation reaction is added, the spins and orbital angular momentum
remain coplanar as the BBHs inspiral, although $\theta_1$ and
$\theta_2$ slowly change on the inspiral timescale.  Not only do
resonant configurations remain resonant, but configurations near
resonance can be captured into resonance during the inspiral.  The
resonances are thus very important for understanding the evolution of
generic BBH systems, although the resonances themselves only occupy a
small portion of the 7-dimensional parameter space characterizing
generic mergers.  We shall review these spin-orbit resonances in more
detail in Sec.~\ref{S:res}.

Bogdanovi\'{c} {\it et al.} \cite{Bogdanovic:2007hp} briefly considered
whether spin-orbit resonances could effectively align SBH spins with the
orbital angular momentum following the merger of gas-poor galaxies.  They
found that for a mass ratio $q = 9/11$ and maximal spins $\chi_1 = \chi_2 =
1$, an isotropic distribution of spins at $r_i = 1000 M$ remains isotropically
distributed when evolved to $r_f = 10 M$.  They therefore concluded that an
alternative mechanism, such as the accretion torques considered later in their
paper, is needed to align the BBH spins with ${\bf L}$.  This conclusion is
supported by a much larger set of PN inspirals presented by Herrmann {\it et
  al.} \cite{Herrmann:2009mr} who found that for equal-mass BBHs, an isotropic
distribution of spins at $40 M$ yields a flat distribution in
$\cos \theta_{12}$ at $7.4 M$.  Here and in this paper $\theta_{12}$ is the
angle between the two spins $\xa$ and $\xb$.  In the final plot of their
paper, Herrmann {\it et al.}  \cite{Herrmann:2009mr} revealed their discovery
of an anti-correlation between the initial and final values of
$\cos \theta_{12}$ for $q = 2/3$ BBHs with equal dimensionless spins $\chi_1 =
\chi_2 = 0.05$.  Investigation of this anti-correlation was left to future
work.  Lousto {\it et al.}  \cite{Lousto:2009ka} also found indications that
an initially isotropic distribution of spins can become non-isotropic during
the PN stage of the inspiral.  For a range of mass ratios $1/16 \leq q \leq 1$
and equal spins $\chi_1 = \chi_2 = (0.485, 0.686, 0.97)$, they found that an
isotropic spin distribution at $50 M$ develops a slight but statistically
significant tendency towards anti-alignment with the orbital angular momentum
${\bf L}$.  This amplitude of anti-alignment scales linearly in the BBH spin
magnitudes and appears to decrease as $q \to 0$.

We perform our own study of PN spin evolution from $r_i$ to $r_f$ for several
reasons.  BBHs get locked into spin-orbit resonances at a separation
\begin{equation} \label{E:rlock}
r_{\rm lock} \propto \left( \frac{\chi_1 \cos \theta_1 - q^2 \chi_2
\cos \theta_2 }{1 - q^2} \right)^2 M~,
\end{equation}
which can become large in the equal-mass $(q \to 1)$ limit
\cite{Schnittman:2004vq}.  This limit is important, as the largest recoil
velocities occur for nearly equal-mass mergers.  Numerical integration of the
PN equations has shown that for a mass ratio $q = 9/11$, spin-orbit resonances
affect spin orientations at separations $r \simeq 1000 M$.  This is a much
larger separation than was considered in previous studies
\cite{Herrmann:2009mr,Lousto:2009ka} of spin alignment, which may therefore
have failed to capture the full magnitude of the effect. These studies also
focused on whether an initially isotropic distribution of spins becomes
anisotropic just prior to merger.  However, as discussed above, tidal torques
from a circumbinary disk partially align spins with the orbital angular
momentum at separations $r \gg r_{\rm GW}$ before relativistic effects become
important.  As we will show in Sec.~\ref{S:align}, such partially aligned
distributions can be strongly affected by spin-orbit resonances despite the
fact that isotropic distributions remain nearly isotropic.  We will consider
how spin precession affects the final spin magnitudes and directions in
Sec.~\ref{S:dist}.  The evolution of the distribution of BBH spin directions
between $r_i$ and $r_f$ changes the distribution of final spin magnitudes and
directions from what it would have been in the absence of precession.  In
addition, spin precession introduces a fundamental uncertainty in predicting
the final spin of a given BBH system.  At large separations, a small
uncertainty in the separation leads to an uncertainty in the predicted time
until merger that exceeds the precession time.  In this case, one cannot
predict at what phase of the spin precession the merger will occur and thus
the resulting final spin.  We will explore this uncertainty in
Sec.~\ref{S:err}.  A brief discussion of the chief findings of this paper is
given in Sec.~\ref{S:disc}.

\section{Post-Newtonian Evolution} \label{S:PN}

We evolve spinning BBH systems along a sequence of quasi-circular
orbits according to the PN equations of motion for precessing binaries
first derived by Kidder \cite{Kidder:1995zr}, and later used by
Buonanno, Chen and Vallisneri to build matched-filtering template
families for GW detection \cite{Buonanno:2002fy}. The adiabatic
evolution of the binary's orbital frequency is described including
terms up to 3.5PN order, and spin effects are included up to 2PN
order. These evolution equations were chosen for consistency with
previous work, in particular with the study by Barausse and Rezzolla
\cite{Barausse:2009uz} of the final spin resulting from the
coalescence of BBHs and with the statistical investigation of spinning
BBH evolutions using Graphics Processing Units by Herrmann {\it et
al.} \cite{Herrmann:2009mr}.  Lousto {\it et al.} \cite{Lousto:2009ka}
evolved a large sample of spinning BBH systems using a non-resummed,
PN expanded Hamiltonian. The convergence properties of non-resummed
Hamiltonians for spinning BBH systems are somewhat problematic (see
e.g. Fig.~1 of Ref.~\cite{Buonanno:2005xu}), and it will be
interesting to repeat these statistical investigations of precessing
BBH systems using the effective-one-body resummations of the PN
Hamiltonian recently proposed by Barausse {\it et al.}
\cite{Barausse:2009aa,Barausse:2009xi}.

In our simulations, the spins evolve according to
\begin{subequations} \label{E:SP}
  \begin{eqnarray} \label{E:SP1}
	\dSa &=& \oaSPFa \times \Sa \, , \\ \label{E:SP2}
	\dSb &=& \oaSPFb \times \Sb \, ,
  \end{eqnarray}
\end{subequations}
where
\begin{subequations} \label{E:oaP}
  \begin{eqnarray} 
\lefteqn{\oaSPFa = } \label{E:oa1}\\
   && \frac{1}{2r^3}\left[ \left( 4 + 3q
      - \frac{3(\Sb+q\Sa) \cdot \LN}{L_{N}^2} \right) \LN + \Sb
      \right]\,,\nn
      \\ 
\lefteqn{\oaSPFb = } \label{E:oa2}\\
   &&  \frac{1}{2r^3}\left[ \left( 4 + \frac{3}{q}
       - \frac{3(\Sa+q^{-1}\Sb) \cdot \LN}{L_{N}^2} \right) \LN + \Sa 
       \right]\nn
  \end{eqnarray}
\end{subequations}
are the spin precession frequencies averaged over a circular orbit, including
the quadrupole-monopole interaction \cite{Racine:2008qv},
\begin{equation} \label{E:LNewt}
\LN = \eta M \mathbf{r} \times \mathbf{v} = \frac{\eta M^2}{(M \omega)^{1/3}} \hLN
\end{equation}
is the Newtonian orbital angular momentum, and
\be
\omega = \left( \frac{M}{r^3} \right)^{1/2}
\ee
is the orbital frequency.  In the absence of gravitational radiation,
$\mathbf{J}$ and $|\LN|$ are constant, implying that the direction of the
orbital angular momentum evolves according to
\be \label{E:hLNdot}
\dhLN=-\f{(M\omega)^{1/3}}{\eta M^2}\f{d\vS}{dt}
\ee
where $\vS=\Sa+\Sb$.  Once radiation reaction is included, the orbital
frequency slowly evolves as
\begin{widetext}
\begin{eqnarray}
\label{omegadot}
\dot\omega&=&
\omega^2\f{96}{5}\eta (M\omega)^{5/3}
\left\{
1-\f{743+924\eta}{336}(M\omega)^{2/3}
+\left[
\left(\f{19}{3}\eta-\f{113}{12}\right)\xsum\cdot \hLN
-\f{113\delta}{12} \xdiff \cdot \hLN+4\pi
\right]\right.(M\omega)\\
&+&
\Bigl\{
\left(
\f{34103}{18144}+\f{13661}{2016}\eta+\f{59}{18}\eta^2
\right)
-\f{\eta\chi_1\chi_2}{48}
\left(247\hSa\cdot\hSb-721(\hLN\cdot \hSa)(\hLN\cdot \hSb)\right)
\nn\\
&+&
\sum_{i=1}^2\frac{(m_i\chi_i)^2}{M^2}
\left[ \frac{5}{2}\left( 3(\hLN\cdot \hSi)^2-1 \right)
+
\frac{1}{96}\left( 7-(\hLN\cdot \hSi)^2 \right) \right]
\Bigr\}
(M\omega)^{4/3}
\nn\\
&-&\f{4159+15876\eta}{672}\pi(M\omega)^{5/3}
+\left[
\left(
\f{16447322263}{139708800}-\f{1712\gamma_E}{105}
+\f{16\pi^2}{3}
\right)+\left(
-\f{273811877}{1088640}+\f{451\pi^2}{48}
-\f{88}{3}\hat \theta \eta
\right)\eta
\right.\nn\\
&+&\f{541}{896}\eta^2
-\f{5605}{2592}\eta^3
-\left.
\left.
\f{856}{105}\log[16(M\omega)^{2/3}]
\right](M\omega)^2
+\left(
-\f{4415}{4032}+\f{358675}{6048}\eta+\f{91495}{1512}\eta^2
\right)\pi
(M\omega)^{7/3}
\right\}
\nn
\end{eqnarray}
\end{widetext}
where $\gamma_E \simeq 0.577$ is Euler's constant, $\hat\theta \equiv
1039/4620$, and we have defined
\begin{subequations}
\begin{eqnarray}
\xsum&\equiv&\f{1}{2}(\xa+\xb)\,,\\
\xdiff&\equiv&\f{1}{2}(\xa-\xb)\,.
\end{eqnarray}
\end{subequations}
The two terms in square parentheses on the third line of Eq.~(\ref{omegadot})
are due to the quadrupole-monopole interaction \cite{Poisson:1997ha} and to
the spin-spin self interaction \cite{Mikoczi:2005dn}, respectively, and they
were neglected in the statistical study of Ref.~\cite{Herrmann:2009mr}. Their
sum agrees with Eq.~(5.17) of Ref.~\cite{Racine:2008kj}.

The numerical integration of this system of ordinary differential
equations is performed using the adaptive stepsize integrator {\sc
StepperDopr5} \cite{NR}. The evolution of any given BBH system is
specified by the following parameters: the initial orbital frequency
$\omega_i$, the binary's mass ratio $q \equiv m_2/m_1$, the
dimensionless magnitude of each spin $\chi_i$, and the relative
orientation $(\theta_i,\phi_i)$ of each spin with respect to the
orbital angular momentum at time $t=0$ ($i=1,\,2$).  To monitor the
variables along the whole evolution we output all quantities using a
constant logarithmic spacing in the orbital frequency at low
frequencies, and the stepsize as used in the integrator at high
frequencies.  Typically this results in a total of about $64,000$
points in the range $M\omega\in [M\omega_i,\,M\omega_f]$, where
$M\omega_i=3.16\times 10^{-5}$ and $M\omega_f=0.1$. Numerical
experimentation indicates that a tolerance parameter {\sc
atol}$=2\times 10^{-8}$ in the adaptive stepsize integrator is
sufficient for a pointwise accuracy of order $1\%$ or better in the
final quantities. Therefore the error induced by the numerical
integrations of the PN equations of motion is subdominant with respect
to the errors induced by precessional effects and by fits of the
numerical simulations, which will be one of the main topics of this
paper.

\section{Spin-orbit Resonances} \label{S:res}

In this Section, we review the equilibrium configurations of BBH spins first
presented in Schnittman \cite{Schnittman:2004vq} for which the Newtonian
orbital angular momentum $\LN$ and individual spins $\mathbf{S}_{1,2}$ all
precess at the same resonant frequency.  As discussed briefly in the
Introduction, at a given binary separation $r$ fully general quasi-circular
BBHs are described by 7 parameters: the mass ratio $q$ and the 3 components of
each black hole spin.  In spherical coordinates with $\LN$ defining the
$z$-axis, each spin is given by its magnitude $S_{i} = m_{i}^2 \chi_{i}$ and
direction $(\theta_{i},\,\phi_{i})$ $(i=1,\,2)$.  In the PN limit for which
this analysis is valid, a clear hierarchy
\begin{equation} \label{E:thier}
t_{\rm orb} \ll t_p \ll t_{\rm GW}
\end{equation}
exists between the orbital time $t_{\rm orb} \propto r^{3/2}$, the precession
time $t_p \sim \Omega_{1,2}^{-1} \propto r^{5/2}$, and the radiation time
$t_{\rm GW} \sim \dot{E}_{\rm GW}/E \propto r^4$.  This hierarchy implies that
the BBH spins will precess many times before merger leaving only their {\it
  relative} angular separation $\DP \equiv \phi_2 - \phi_1$ in the orbital
plane well defined.  This reduces the BBH parameter space to 6 dimensions.
Since the mass ratio and individual spin magnitudes are preserved during the
inspiral, a given BBH evolves through the 3-dimensional parameter space
$(\theta_1, \theta_2, \DP)$ on the precession timescale $t_p$.  This
evolution is governed by the spin precession equations (\ref{E:SP}).

Schnittman \cite{Schnittman:2004vq} discovered a one-parameter family of
equilibrium solutions to these equations for which $(\theta_1, \theta_2, \DP)$
remain fixed on the precession timescale $t_p$.  These solutions have $\DP =
0^\circ$ or $180^\circ$, implying that $\LN$, $\Sa$ and $\Sb$ all lie in a
plane and precess at the same resonant frequency about the total angular
momentum $\mathbf{J}$, which remains fixed in the absence of gravitational
radiation.  The values of $\theta_{1,2}$ for these resonances can be
determined by requiring the first and second time derivatives of $\Sa \cdot
\Sb$ to vanish.  This is equivalent to satisfying the algebraic constraint
\begin{multline} \label{E:rescon}
(\oaSPFa \times \Sa) \cdot [\Sb \times (\LN + \Sa)] \\
= (\oaSPFb \times \Sb) \cdot [\Sa \times (\LN + \Sb)]~.
\end{multline}
Since $\LN$ appears in Eq.~(\ref{E:rescon}) both explicitly and implicitly
through $\bar{\boldsymbol{\Omega}}_{1,2}$, the resonant values of
$\theta_{1,2}$ vary with the binary separation.  This is crucial, as otherwise
these one-parameter families of resonances would affect only a small portion
of the 3-dimensional parameter space $(\theta_1, \theta_2, \DP)$ through which
generic BBH configurations evolve.  As gravitational radiation slowly extracts
angular momentum from the binary on the radiation time $t_{\rm GW}$, the
resonances sweep through a significant portion of the $(\theta_1, \theta_2)$
plane.  The angular separation $\DP$ of a generic BBH is varying on the much
shorter precession time $t_p$, and thus has a significant chance to closely
approach the resonant values $\DP = 0^\circ$ or $180^\circ$ at some point
during the long inspiral.  Such generic BBHs will be strongly influenced or
even captured by the spin-orbit resonances, as we will see in detail in
Sec.~\ref{S:align}.

\begin{figure}[t!]
\begin{center}
\includegraphics[width=3.5in]{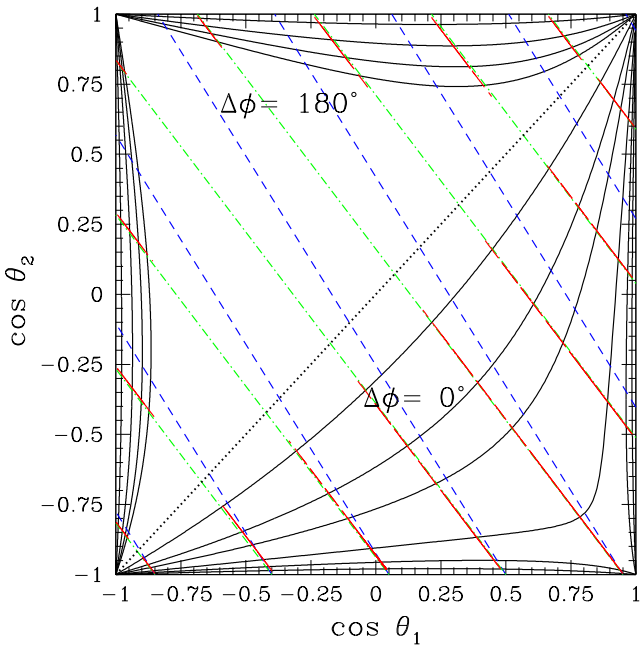}
\end{center}
\caption{Spin-orbit resonances for maximally spinning BBHs with a mass ratio
  of $q = 9/11$.  The dotted black diagonal indicates where $\theta_1
  = \theta_2$.  Solid black curves below (above) this diagonal show
  $(\theta_1, \theta_2)$ for the one-parameter families of equilibrium
  spin configurations with $\DP = 0^\circ (180^\circ)$ at different
  fixed binary separations.  Approaching the diagonal from below,
  these curves correspond to separations $r = 1000, 500, 250, 100, 50,
  10 M$.  The curves approaching from above correspond to separations
  $r = 250, 50, 20, 10 M$.  The long-dashed red curves show how
  $\theta_{1,2}$ evolve as members of these resonant families inspiral
  from $r_i = 1000 M$ to $r_f = 10 M$.  The projection $\mathbf{S}
  \cdot \hLN$ of the total spin $\mathbf{S}$ onto the orbital angular
  momentum $\LN$ is constant along the short-dashed blue lines, while
  the projection $\mathbf{S}_0 \cdot \hLN$ of the EOB spin
  $\mathbf{S}_0$ is constant along the dot-dashed green lines.}
  \label{F:resq1.22}
\end{figure}

\begin{figure}[t!]
\begin{center}
\includegraphics[width=3.5in]{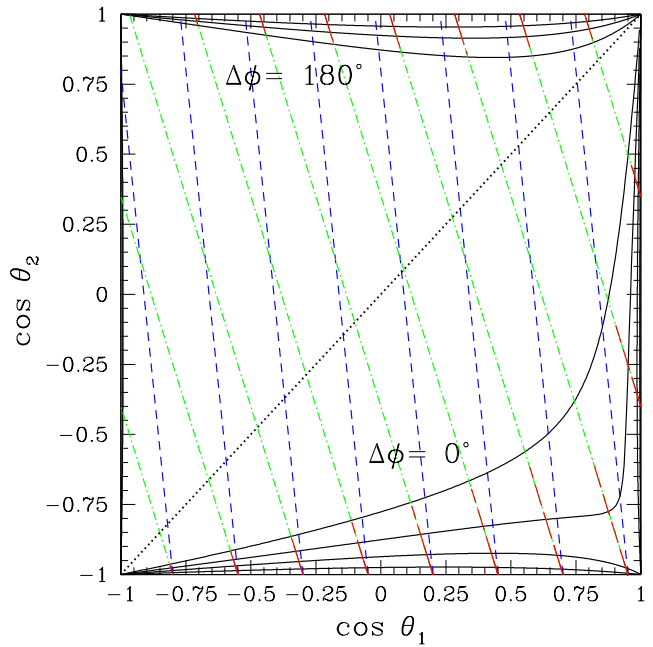}
\end{center}
\caption{Spin-orbit resonances for maximally spinning BBHs with a mass ratio
  of $q = 1/3$.  Other than the different mass ratio, this figure is
  very similar to Fig.~\ref{F:resq1.22}.  The solid black curves
  approaching the diagonal from below correspond to the families of
  resonant spin configurations at $r = 50, 20, 10, 5 M$, while those
  approaching from above correspond to separations $r = 20, 10, 5 M$.}
\label{F:resq3}
\end{figure}

We show the dependence of the spin-orbit resonances on $r$ for maximally
spinning BBHs in Figs.~\ref{F:resq1.22} and \ref{F:resq3}.  Those resonances
with $\DP = 0^\circ$ (shown in Fig.~2 of \cite{Schnittman:2004vq}) always have
$\theta_1 < \theta_2$, and thus appear below the diagonal $\cos \theta_1 =
\cos \theta_2$ in our Figs.~\ref{F:resq1.22} and \ref{F:resq3}.  Those
resonances with $\DP = 180^\circ$ (shown in Fig.~3 of
\cite{Schnittman:2004vq}) have $\theta_1 > \theta_2$ and therefore
appear above the diagonal in our Figs.~\ref{F:resq1.22} and
\ref{F:resq3}.  We plot $(\cos \theta_1, \cos \theta_2)$ rather than
$(\theta_1, \theta_2)$ like \cite{Schnittman:2004vq} because
isotropically oriented spins should have a flat distribution in these
variables.

In the limit $r \to \infty$, so that also $|\LN|\to \infty$, the resonant
configurations either have $\Sa$ or $\Sb$ aligned or anti-aligned with $\LN$
(either $\theta_1$ or $\theta_2$ equals to $0^\circ$ or $180^\circ$).  This
corresponds to the four edges of the plot in Fig.~\ref{F:resq1.22}.  For
smaller fixed values of $|\LN|$, the values $(\theta_1, \theta_2)$ for the
one-parameter families of resonant configurations approach the diagonal
$\theta_1 = \theta_2$.  BBHs in spin-orbit resonances at large values of
$|\LN|$ (large $r$) remain resonant as they inspiral. As gravitational
radiation carries away angular momentum, $r$ decreases and $\theta_{1,2}$ for
individual resonant BBHs evolves towards this diagonal along the red
long-dashed curves in Fig.~\ref{F:resq1.22}.  For resonances with $\DP =
0^\circ$ (those below the diagonal), this evolution aligns the two spins with
each other.  Symmetry implies that aligning the spins with each other will
lead to larger final spins and smaller recoil velocities
\cite{Boyle:2007sz,Boyle:2007ru}.

The projection
\begin{equation} \label{E:Spar}
\mathbf{S} \cdot \hLN = S_1 \cos \theta_1 + S_2 \cos \theta_2
\end{equation}
of the total spin $\mathbf{S} \equiv \Sa + \Sb$ parallel to the
orbital angular momentum is constant along the short-dashed blue lines
in Figs.~\ref{F:resq1.22} and \ref{F:resq3}.  These blue lines have
steeper slopes than the red lines along which the resonant binaries
inspiral.  This implies that the total spin $\mathbf{S}$ becomes
anti-aligned (aligned) with the orbital angular momentum for resonant
configurations with $\DP = 0^\circ (180^\circ)$, leading to smaller
(larger) final spins.  The families of resonances with $\DP = 0^\circ$
(below the diagonal) sweep through a larger area of the $(\cos
\theta_1, \cos \theta_2)$ plane as the BBHs inspiral, and approach the
diagonal more closely.  This implies that anti-alignment may be more
effective than alignment, which might explain the ``small but
statistically significant bias of the distribution towards
counter-alignment'' in $\mathbf{S} \cdot \hLN$ noted in Lousto {\it et
al.} \cite{Lousto:2009ka}.  However, Table IV of \cite{Lousto:2009ka}
indicates that both $\Sa$ and $\Sb$ individually become anti-aligned
with $\hLN$, whereas the spin-orbit resonances would align one black
hole while anti-aligning the other.  All of the PN evolutions in
Lousto {\it et al.} \cite{Lousto:2009ka} begin at separations of $r =
50 M$, which corresponds to the $\DP = 0^\circ$ curve in
Fig.~\ref{F:resq1.22} that is second closest to the diagonal.  The
resonances sweep through most of the plane below the diagonal at
larger separations, suggesting that these short-duration PN evolutions
may have failed to capture the full magnitude of the anti-alignment.
We will investigate this possibility in Sec.~\ref{S:align}.

Another interesting feature of Figs.~\ref{F:resq1.22} and
\ref{F:resq3} is that the red long-dashed curves along which the BBHs
inspiral are nearly parallel to the dot-dashed green lines along which
the projection $\Szero \cdot \hLN$ of the effective-one-body (EOB) spin
\cite{Damour:2001tu}
\begin{equation} \label{E:S0}
\Szero \equiv (1 + q)\Sa + (1 + q^{-1})\Sb
\end{equation}
is constant.  The conservation of this quantity at 2PN order was first
noted in Ref.~\cite{Racine:2008qv} and follows directly from
Eqs.~(\ref{E:SP}), (\ref{E:oaP}), and (\ref{E:hLNdot}).  The
conservation of $\Szero \cdot \hLN$ rather than $\mathbf{S} \cdot \hLN$
itself allows for the possible alignment of the total spin
$\mathbf{S}$ discussed in the previous paragraph.

We conclude this Section by briefly discussing how the spin-orbit
resonances vary with the mass ratio $q$, as can be seen by comparing
the $q = 9/11$ resonances in Fig.~\ref{F:resq1.22} with the $q = 1/3$
resonances in Fig.~\ref{F:resq3}.  The most pronounced differences are
that the $q = 1/3$ resonances sweep away from the edges of the $(\cos
\theta_1, \cos \theta_2)$ plane at much smaller values of the
separation $r$, and do not approach the diagonal as closely.  This is
consistent with the decreasing value of $r_{\rm lock}$ in
Eq.~(\ref{E:rlock}) as $q \to 0$.  In this limit both $t_p$ and
$t_{\rm GW}$ are proportional to $q^{-1}$, implying that generic BBHs
will be less likely to be affected by the resonances as they sweep
through the plane over a smaller range in $r$.  BBHs already in a
resonant configuration will also be less affected since the resonant
curves do not approach the diagonal as closely.  The red long-dashed
curves showing the inspiral of resonant configurations have steeper
slopes for $q = 1/3$, consistent with the larger black hole being
immune to its smaller companion in the limit $q \to 0$.  This seems to
contradict the puzzling result presented in Table IV of Lousto {\it et
al.}  \cite{Lousto:2009ka} that it is the {\it smaller} companion that
remains randomly distributed during the inspiral.  We will examine
this behavior as well in the next Section.

\section{Spin Alignment} \label{S:align}

\begin{figure}[t!]
\begin{center}
\includegraphics[width=3.5in]{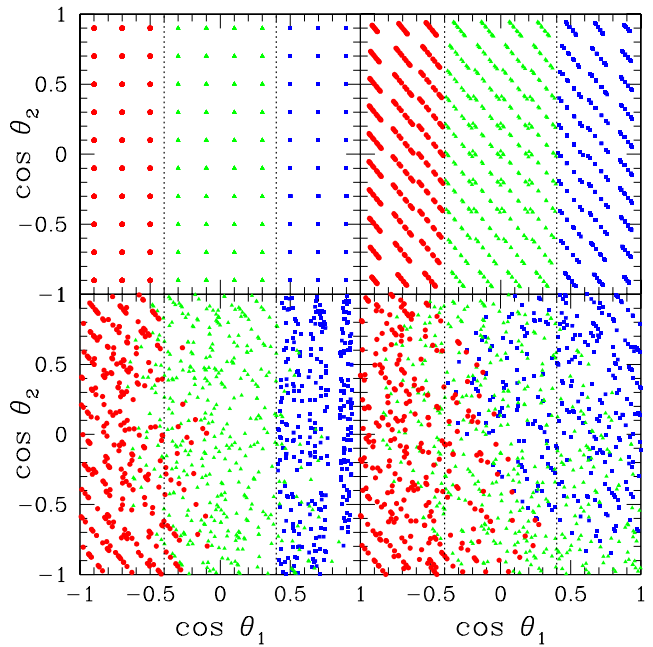}
\end{center}
\caption{Distributions of $(\cos \theta_1, \cos \theta_2)$ at different
  separations $r$ for 1000 initially isotropic maximally spinning BBHs with a
  mass ratio $q = 9/11$.  The top left panel shows the initial $10 \times 10
  \times 10$ grid, evenly spaced in $(\cos \theta_1, \cos \theta_2, \Delta
  \phi)$.  The dotted vertical lines show $\cos \theta_1 = \pm 0.4$.  The 300
  blue squares initially have $\cos \theta_1 > 0.4$, the 400 green triangles
  initially have $-0.4 < \cos \theta_1 < 0.4$, and the 300 red circles
  initially have $\cos \theta_1 < -0.4$.  The values of $(\theta_1, \theta_2)$
  for these BBHs are shown in the top right, bottom left, and bottom right
  panels after they have inspiraled to separations of $r = 1000, 100$ and $10
  M$ respectively.} \label{F:t1t2}
\end{figure}

\begin{figure}[t!]
\begin{center}
\includegraphics[width=3.5in]{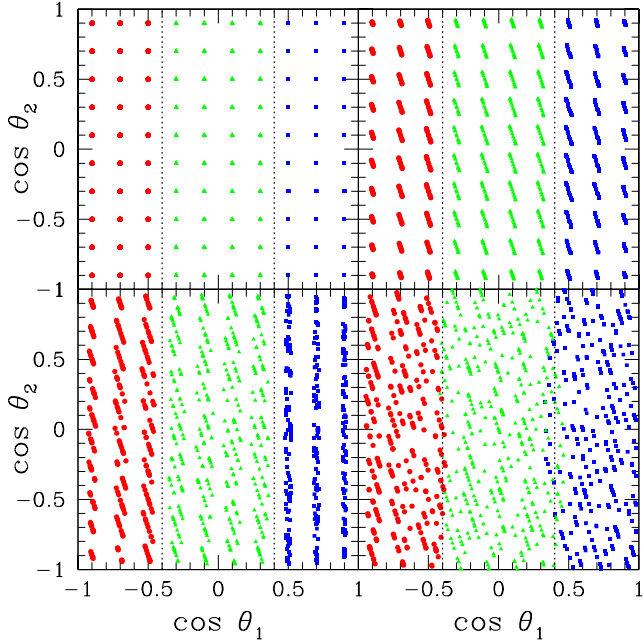}
\end{center}
\caption{Distributions of $(\cos \theta_1, \cos \theta_2)$ at different
  separations $r$ for 1000 initially isotropic maximally spinning BBHs with a
  mass ratio $q = 1/3$.  The different panels, points, and lines are the same
  as those given for $q = 9/11$ in Fig.~\ref{F:t1t2}.}
\label{F:q3t1t2}
\end{figure}

In this Section, we examine the extent to which the spins of {\it
generic} (i.e. misaligned) BBH configurations become aligned with the
orbital angular momentum and each other as the BBHs inspiral from $r_i
= 1000 M$ to $r_f = 10 M$.  Although we use maximally spinning BBHs to
demonstrate this alignment, the magnitude of the alignment is
comparable for all BBHs with $\chi_{1,2} \gtrsim 0.5$ as shown in
Fig.~11 of \cite{Schnittman:2004vq}.  We first consider initial spin
configurations given by a uniform $10 \times 10 \times 10$ grid evenly
spaced in $(\cos \theta_1, \cos \theta_2, \Delta \phi)$.  This
distribution is isotropic, and would be expected in the absence of an
astrophysical mechanism to align the spins.  BBHs with isotropically
oriented spins might form in gas-poor mergers of SBHs and mergers of
stellar-mass black holes in dense clusters.

In Fig.~\ref{F:t1t2}, we show how the distribution of $(\cos \theta_1,
\cos \theta_2)$ evolves as maximally spinning BBHs with a mass ratio $q = 9/11$
inspiral from slightly beyond $r_i = 1000 M$ to $r_f = 10 M$.  The top
left panel shows our initial evenly spaced $10 \times 10 \times 10$
grid.  The points are colored to indicate their {\it initial} value of
$\cos \theta_1$: blue squares begin with $\cos \theta_1 > 0.4$
($\theta_1\lesssim 66^\circ$), green triangles with $-0.4 < \cos
\theta_1 < 0.4$, and red circles with $\cos \theta_1 < -0.4$.  The dotted
vertical lines $\cos \theta = \pm 0.4$ denote these boundaries.  Only
100 points are visible in the top left panel, as the different values
of $\DP$ cannot be distinguished in this two-dimensional projection.
Spin precession reveals all 1000 points after the BBHs have inspiraled
to $r_i = 1000 M$ as seen in the top right panel.  Notice that the
spins of all 1000 BBHs precess in a way that conserves the projection
of $\Szero$ onto $\hLN$ (parallel to the dot-dashed green lines in
Fig.~\ref{F:resq1.22}).  This is not a special feature of the
spin-orbit resonances, but occurs for generically oriented spins as
well.  These generic spin configurations do not individually preserve
$(\cos \theta_1, \cos \theta_2)$ over a precession time $t_p$ like the
resonant configurations do, but they do preserve the combination
$\Szero \cdot \hLN$.  This precession continues as the BBHs inspiral to
$r = 100 M$ and $r_f = 10 M$ as shown in the bottom left and bottom
right panels of Fig.~\ref{F:t1t2}.  By the time they reach $r_f = 10
M$ the green points have diffused to fill most of the $(\cos
\theta_1, \cos \theta_2)$ plane, while the blue (red) points have diffused
into the upper right (lower left) portion of the middle $-0.4 < \cos
\theta_1 < 0.4$ region.  The bottom right panel, if the points had not
been colored, would reproduce Fig.~1 of Bogdanovic {\it et al.}
\cite{Bogdanovic:2007hp} and therefore support their conclusion that
isotropically distributed spins remain isotropic as they inspiral.
However, the colors reveal that PN evolution can drastically alter
spin distributions that have been partially aligned by a circumbinary
disk.  For example, if the spin of the more massive black hole was
aligned so that $\cos \theta_1 > 0.4$ at $r_i = 1000 M$ (shown by our
blue points), by the time the binary reached $r_f = 10 M$ the larger
spin could easily lie in the orbital plane and thus give rise to a
smaller final spin and potentially large ``superkick''
\cite{Gonzalez:2007hi,Campanelli:2007ew}.

For comparison, we show the inspiral of the same $10 \times 10 \times
10$ grid of maximally spinning BBHs with a mass ratio $q = 1/3$ in
Fig.~\ref{F:q3t1t2}.  The points diffuse along the steeper lines that
preserve $\Szero \cdot \hLN$ for this less equal mass ratio.  This
inhibits their ability to diffuse across the $\cos \theta_1 = \pm 0.4$
boundaries, again shown by the vertical dotted lines.  Even at $r_f =
10 M$ only a few points have trickled between the three regions.
Since the spin of the more massive black hole remains aligned with the
orbital angular momentum, one would expect a large final spin and an
absence of superkicks for such small mass ratios.  We will examine in
detail how spin alignment affects recoil-velocity distributions in
future work.

\begin{figure}[t!]
\begin{center}
\includegraphics[width=3.5in]{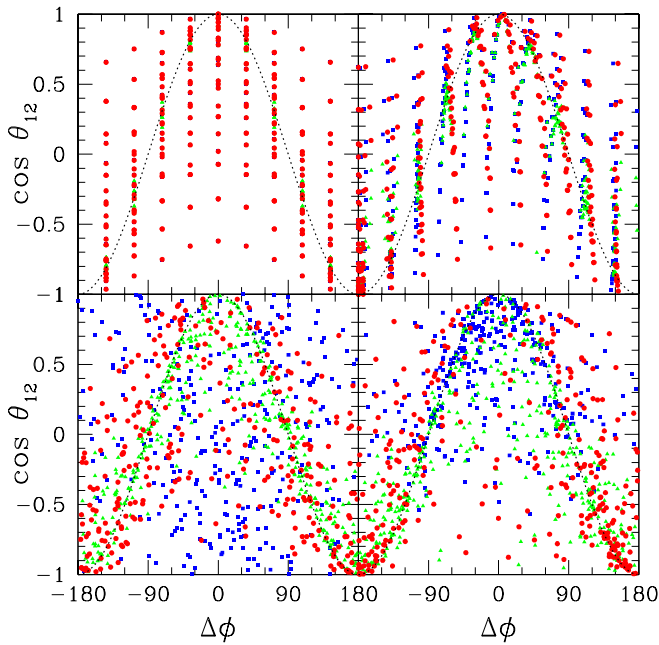}
\end{center}
\caption{Distributions of $(\Delta \phi, \cos \theta_{12})$ at different
  separations $r$ for 1000 initially isotropic maximally spinning BBHs with a
  mass ratio $q = 9/11$.  The top left panel shows the initial $10 \times 10
  \times 10$ grid of BBH spin configurations, evenly spaced in $(\cos
  \theta_1, \cos \theta_2, \Delta \phi)$.  This distribution is peaked about
  the curve $\cos \theta_{12} = \cos \Delta \phi$ shown by the dotted curve.
  The points are colored according to their initial values of $\cos \theta_1$
  as in Fig.~\ref{F:t1t2}.  The top right, bottom left, and bottom right
  panels show the distribution evolves after the BBHs have inspiraled to $r =
  1000, 100$ and $10 M$ respectively, also as in Fig.~\ref{F:t1t2}.}
\label{F:DPt12}
\end{figure}

\begin{figure}[t!]
\begin{center}
\includegraphics[width=3.5in]{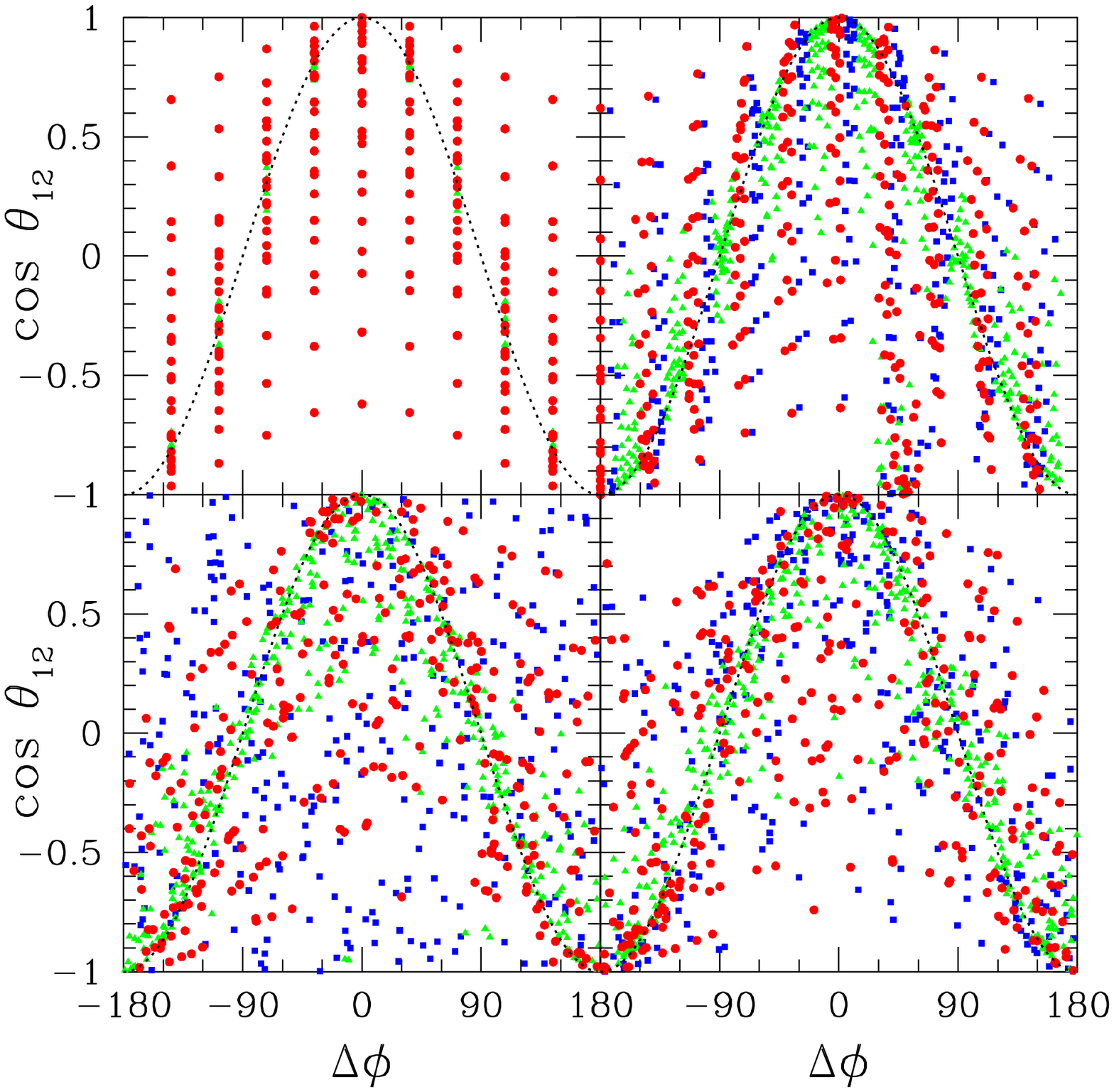}
\end{center}
\caption{Distributions of $(\Delta \phi, \cos \theta_{12})$ at different
  separations $r$ for 1000 initially isotropic maximally spinning BBHs with a
  mass ratio $q = 1/3$.  The different panels, points, and lines are the same
  as those given for $q = 9/11$ in Fig.~\ref{F:DPt12}.}
\label{F:q3DPt12}
\end{figure}

\begin{figure}[t!]
\begin{center}
\includegraphics[width=3.5in]{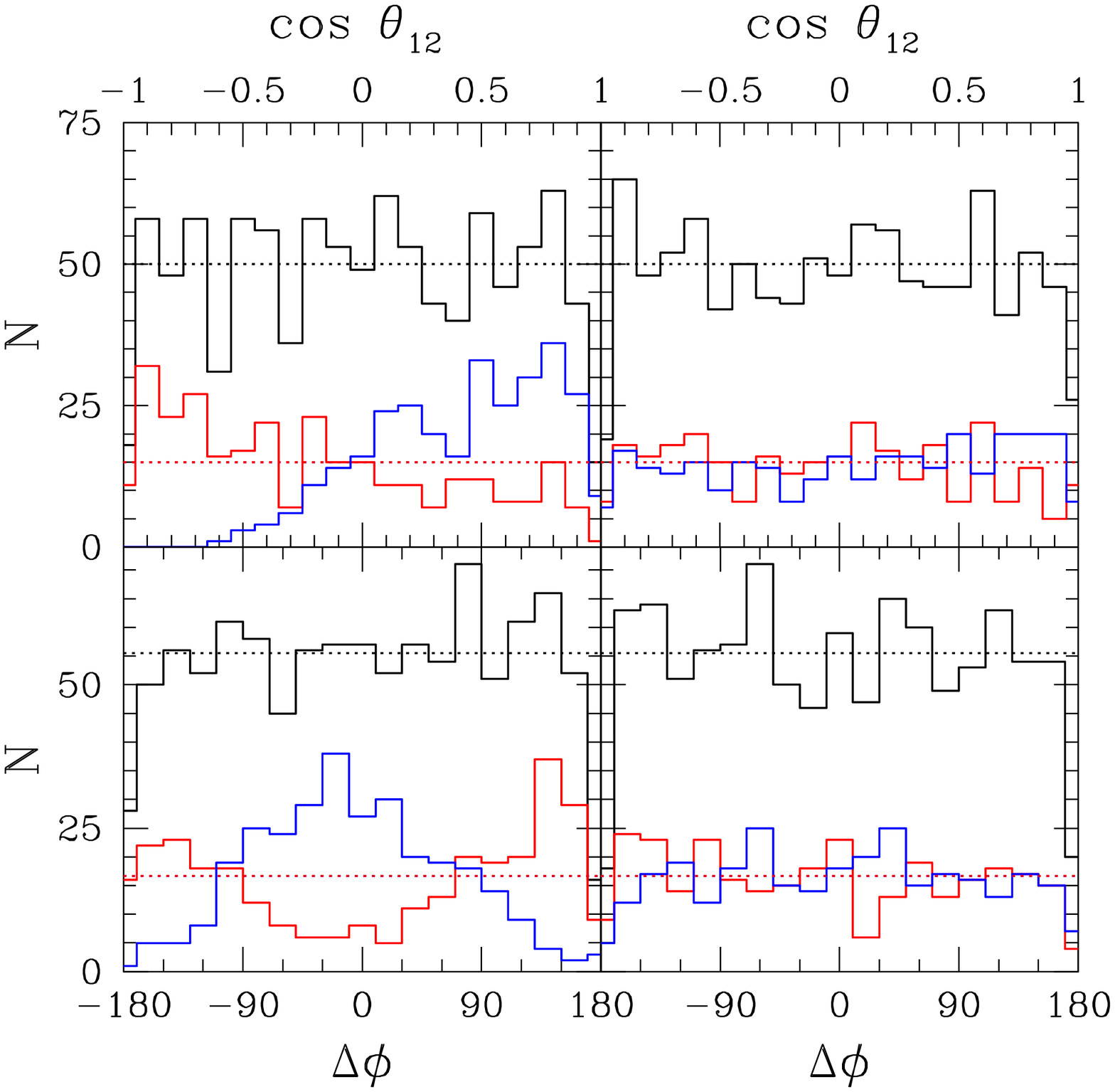}
\end{center}
\caption{Histograms of $\cos \theta_{12}$ and $\DP$ for BBHs with
initially isotropic spins.  The two left panels are for the mass ratio
$q = 9/11$, while the two right panels are for $q = 1/3$.  The two top
panels give the distribution of $\cos \theta_{12}$, while the two
bottom panels give the distribution of $\DP$.  The black curves are
for all 1000 BBHs in the $10 \times 10 \times 10$ grid discussed in
the text, while the blue (red) curves correspond to the blue (red)
points in Figs.~\ref{F:t1t2}-\ref{F:q3DPt12} with initial values $\cos
\theta_1 > 0.4$ $(\cos \theta_1 < -0.4)$.  The horizontal dotted lines
show the initially flat distributions, while the solid lines show the
distributions at $r = 10 M$.} \label{F:pdft12DP}
\end{figure}

In Fig.~\ref{F:DPt12} we show how the joint probability distribution
function for $\Delta \phi$ and $\cos \theta_{12}$ evolves for our
evenly spaced $10 \times 10 \times 10$ grid of initially isotropic BBH
spin configurations.  As defined in the Introduction, $\cos
\theta_{12}$ is the cosine of the angle between $\Sa$ and $\Sb$.  It
can be expressed in terms of the individual spin angles as
\be \label{E:cos12}
\cos \theta_{12} = \sin \theta_1 \sin \theta_2
\cos \Delta \phi + \cos \theta_1 \cos \theta_2~,
\ee
and has a flat distribution between -1 and 1 for isotropic,
uncorrelated spins such as those given by our $10 \times 10 \times 10$
grid.  However, as seen in Eq.~(\ref{E:cos12}), the values of $\cos
\theta_{12}$ and $\cos \Delta \phi$ are correlated; for a given value
of $\Delta \phi$ the distribution of $\cos \theta_{12}$ is peaked
about $\cos \Delta \phi$ for flat distributions of $\cos \theta_1$ and
$\cos \theta_2$.  This can be seen in Fig.~\ref{F:DPt12} from the
clustering of points about the curve $\cos \theta_{12} = \cos \Delta
\phi$.  Although $\cos \theta_{12}$ and $\cos \Delta \phi$ are correlated
even for isotropic spins, geometry implies that both are initially
uncorrelated with the value of $\cos \theta_1$.  This is revealed by
the identical distributions of the red, green, and blue points in the
top left panel of Fig.~\ref{F:DPt12} to within the resolution of our
grid.  These distributions do not remain identical as the BBHs
inspiral from $r_i = 1000 M$ to $r_f = 10 M$.  Influenced by the
$\Delta \phi = 0^\circ$ spin-orbit resonances below the diagonal in
Fig.~\ref{F:resq1.22}, the blue points become concentrated about
$\Delta \phi = 0^\circ, \cos \theta_{12} = 1$ by the time they reach
$r_f$.  The red points, similarly influenced by the $\Delta \phi =
\pm 180^\circ$ resonances above the diagonal in Fig.~\ref{F:resq1.22},
become concentrated about $\Delta \phi = \pm 180^\circ, \cos
\theta_{12} = -1$.  The effect of this spin alignment on the spin of
the final black hole will be explored in detail in the next Section,
while the effect on recoil velocities will be examined in future work.
Qualitatively, alignment of the spins with each other ($\cos
\theta_{12} \to 1$) increases the final spin and reduces the recoil
velocity, while anti-alignment ($\cos \theta_{12} \to -1$) does the
opposite.

The magnitude of this spin alignment is greatly reduced for smaller
mass ratios as seen in Fig.~\ref{F:q3DPt12} for the case $q = 1/3$.
Although the clustering of all the points about $\cos \theta_{12} =
\cos \Delta \phi$ is again apparent, the distributions of the red,
green, and blue points remain similar all the way down to $r_f = 10 M$
as seen in the lower right panel.  The weaker influence of the
spin-orbit resonances for $q = 1/3$ follows from the smaller value of
$r_{\rm lock}$ in Eq.~(\ref{E:rlock}), and is similarly reflected by
the smaller fraction of the $(\cos \theta_1, \cos \theta_2)$ plane
occupied by the resonant curves in Fig.~\ref{F:resq3}.

We have provided histograms of $\cos \theta_{12}$ and $\DP$ in
Fig.~\ref{F:pdft12DP} to clarify the differences between
Figs.~\ref{F:DPt12} and \ref{F:q3DPt12}.  We see that the
distributions of $\cos \theta_{12}$ and $\Delta \phi$ are initially
flat for both mass ratios, but evolve considerably for $q = 9/11$
while remaining nearly flat for $q = 1/3$ within the limits set by
Poisson fluctuations.  The open blue (red) curves in the left panels
of Fig.~\ref{F:pdft12DP} clearly show distributions peaked at $\cos
\theta_{12} = 1, \Delta \phi = 0^\circ$ ($\cos \theta_{12} = -1,
\DP = \pm 180^\circ$).  Such trends are barely noticeable in the
right panels.  We will explore the implications of these findings for
the final spins in the next Section.

\section{Final Spin Distributions} \label{S:dist}

Several attempts have been made to predict the final dimensionless
spin $\xf$ of the black hole resulting from a BBH merger. Initial
attempts focused on finding simple phenomenological fitting formulae
for the final spin resulting from non-spinning, unequal-mass BBH
merger simulations
\cite{Buonanno:2006ui,Berti:2007fi,Buonanno:2007pf}. A group at the
Albert Einstein Institute (AEI) developed a fitting formula that
provides the magnitude and direction of $\xf$ in terms of the initial
spins $\xa$, $\xb$ and the mass ratio $q$
\cite{Rezzolla:2007xa,Rezzolla:2007rd,Rezzolla:2007rz}.  They assumed
that the final spin magnitude could be expressed as a polynomial in
$\chi_1$, $\chi_2$, and the symmetric mass ratio $\eta$, then made
some additional assumptions about the symmetries of this polynomial
dependence and how energy and angular momentum are radiated to reduce
the number of terms in their expression.  The coefficients of the
remaining terms were calibrated using numerical-relativity (NR)
simulations of BBH mergers in which the initial spins were either
aligned or anti-aligned with the orbital angular momentum.  We shall
refer to this older AEI formula as ``AEIo''.  A more recent paper
\cite{Barausse:2009uz} by members of this group uses newer NR
simulations to recalibrate their coefficients, and replaces earlier
assumptions with the conjecture that the final spin points in the
direction of the total angular momentum of the initial BBH at {\it
any} separation.  For consistency, this requires the further
assumption that angular momentum is always radiated in the direction
of the total angular momentum.  We shall refer to this newer AEI
formula as ``AEIn''.  An alternative fitting formula was proposed by a
group at Florida Atlantic University (FAU) \cite{Tichy:2008du}.
Following the procedure outlined in \cite{Boyle:2007sz,Boyle:2007ru},
the FAU group performed 10 equal-mass misaligned simulations to
calibrate the coefficients of fitting formulae for the Cartesian
components of $\xf$.  They then made additional assumptions about the
mass-ratio dependence of these formulae, and found good agreement
between their predictions and independent NR simulations with mass
ratios as small as $q = 5/8$.  We shall refer to the formula of this
group as ``FAU''.  The Rochester Institute of Technology (RIT) group
proposed yet another fitting formula during the preparation of this
paper
\cite{Lousto:2009mf}.  This formula includes higher-order terms in the initial
spins that may ultimately be needed to describe future high-accuracy NR
simulations.  However, current simulations are inadequate to calibrate all the
terms appearing in the RIT formula, so we will not consider its predictions in
this paper.

Other groups have predicted final spins by extrapolating analytical
test-particle calculations to finite mass ratios, rather than
calibrating fitting formulae with NR simulations.  Buonanno, Kidder,
and Lehner (BKL) \cite{Buonanno:2007sv} derived a formula for the
final spin by assuming, as is true in the test-particle limit, that
the angular momentum radiated during the inspiral stage of a BBH
merger exceeds that radiated during the plunge and ringdown.  Using
this assumption, they equated the final spin with the total angular
momentum $\mathbf{J} = \mathbf{L}_{\rm ISCO} + \Sa + \Sb$, where
$\mathbf{L}_{\rm ISCO}$ is the orbital angular momentum at the
innermost stable circular orbit (ISCO) of a test particle of mass
$\eta M$ orbiting a black hole of mass $M$ and dimensionless spin
$\xf$ equal to that of the {\it final} black hole.  This
counterintuitive but inspired choice correctly provides $\xf \to
\xa$ in the $q \to 0$ limit and respects the symmetry of BBH mergers
under exchange of the labels of the two black holes.  Though derived
only from test-particle calculations, the BKL formula is remarkably
successful at predicting final spins even for equal-mass BBH mergers.
Kesden \cite{Kesden:2008ga} slightly modified the BKL spin formula to
account for the energy radiated during the inspiral stage of the
merger.  This change makes the formula accurate to linear order in $q$
in the test-particle limit.  It generically increases the magnitude of
the predicted dimensionless final spin by reducing the predicted final
mass $m_f$ below $M$ in the denominator of the expression $\xf =
S_f/m_{f}^2$.  This increase improves the agreement with NR
simulations of non-spinning BBH mergers, but leads to somewhat larger
final spins than the other formulae for mergers of maximally spinning
BBHs, such as those considered in this paper.  The predictions of this
formula are refered to as ``Kes'' in this paper.

We now present the predictions of the spin formulae summarized above
for various distributions of BBH spins that are allowed to inspiral
from $r_i = 1000 M$ to $r_f = 10 M$.

\subsection{Spin Magnitudes} \label{SS:mag}

\begin{figure}[t!]
\begin{center}
\includegraphics[width=3.5in]{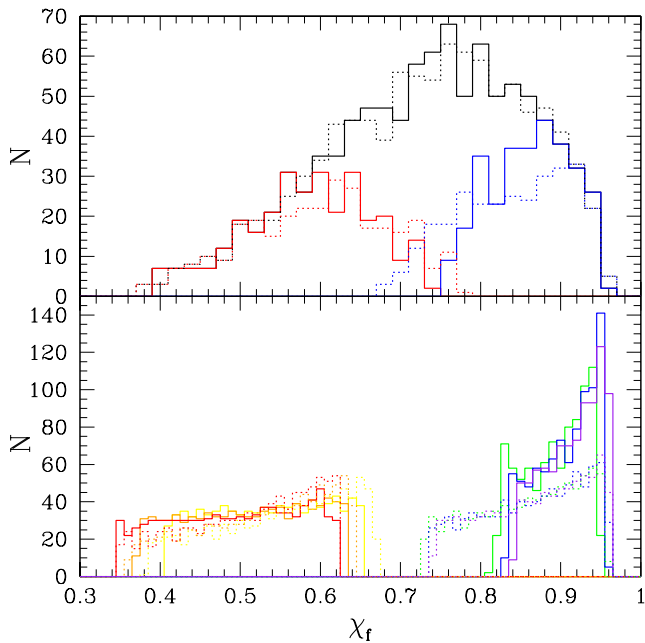}
\end{center}
\caption{{\it Top panel:} Histogram of the final spin $\chi_f$ predicted by
  the AEIn formula for 1000 BBHs with mass ratio $q = 9/11$ and isotropically
  distributed spins at $r_i = 1000 M$.  The blue curves show the subset of 300
  BBHs with the lowest initial values of $\theta_1$, while the red curves show
  the subset of 300 BBHs with the highest initial values of $\theta_1$.  The
  solid curves show the predicted spins if the AEIn formula is applied at $r_f
  = 10 M$ after the BBHs have inspiraled to this separation according to the
  equations of Sec.~\ref{S:PN}.  The dotted curves show the predicted spins if
  the AEIn formula is applied to the initial distribution at $r_i = 1000
  M$. {\it Bottom panel:} Histograms of the predicted final spins for 6 sets
  of BBH mergers with $q = 9/11$, and flat distributions in $\cos \theta_2$
  and $\Delta \phi$ at $r_i = 1000 M$. The red, orange, yellow, green, blue,
  and purple curves have $\theta_1 = 170^\circ, 160^\circ, 150^\circ,
  30^\circ, 20^\circ,$ and $10^\circ$ respectively.  As in the top panel the
  final spins predicted by applying the AEIn formula at $r_i = 1000 M$ are
  shown by dotted curves, while allowing the BBHs to inspiral to $r_f = 10 M$
  before applying the formula leads to the spins shown by the solid
  curves.} \label{F:mag1.22}
\end{figure}

In the top panel of Fig.~\ref{F:mag1.22}, we show the final spin magnitude
$\chi_f$ predicted by the AEIn formula for the evenly spaced $10 \times 10
\times 10$ grid of maximally spinning BBHs with $q = 9/11$ described in
Sec.~\ref{S:align}.  The other spin formulae give very similar results; the
mean and variance of the final spin distributions predicted by the other
formulae for some of the initial distributions described below are provided in
Table~\ref{T:fspin}.  As in Figs.~\ref{F:t1t2}-\ref{F:pdft12DP}, the black
curves in Fig.~\ref{F:mag1.22} refer to all 1000 BBHs, the blue curves to the
subset of 300 BBHs with the lowest values of $\theta_1$, and the red curves to
the subset of 300 BBHs with the highest values of $\theta_1$.  The dotted
curves give the final spin distribution predicted for the BBH spin
configurations at their initial separation $r_i = 1000 M$, while the solid
curves give the final spin distribution predicted when these {\it same} BBHs
are allowed to inspiral to $r_f = 10 M$ according to the PN evolution
described in Sec.~\ref{S:PN}.  The AEIn formula is unique in that it claims to
accurately predict final spins at all separations; separations as large as $r
= 2 \times 10^4 M$ were considered in \cite{Barausse:2009uz}.  The other
fitting formulae were intended to apply at $r_{\rm NR} \simeq 10 M$, the
starting point for the NR simulations with which their coefficients were
calibrated.  The BKL and Kes formulae were designed for use at the ISCO.
Although strictly speaking the formulae other than AEIn cannot be applied to
widely separated BBHs, one can imagine that the BBHs inspiral to $r_f = 10 M$
without spin precession where these formulae are valid.  It is in this sense
that we consider the predictions of these other formulae when we claim in this
Section to apply them to BBH spin configuration at $r_i = 1000 M$.

\begin{figure}[t!]
\begin{center}
\includegraphics[width=3.5in]{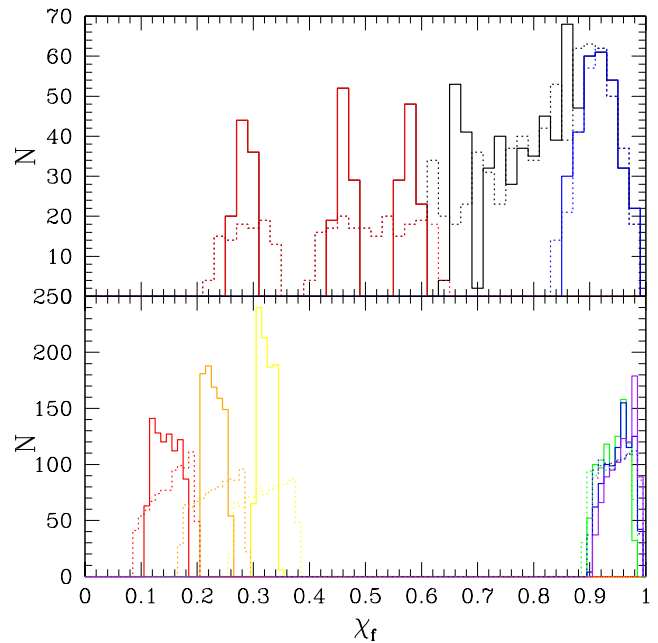}
\end{center}
\caption{Histograms of the final spins $\chi_f$ predicted by the AEIn
  formula for the same sets of BBHs presented in Fig.~\ref{F:mag1.22},
  but with the mass ratio $q = 1/3$ instead of $q = 9/11$.  As in that
  figure, the predictions made at $r_i = 1000 M$ are shown with dotted
  curves, those made at $r_f = 10 M$ are shown with solid curves.  The
  black curves in the top panel show the full set of 1000 BBHs, while
  the blue (red) curves show the subset of 300 BBHs with the lowest
  (highest) initial values of $\theta_1$.  In the lower panel, the
  red, orange, and yellow curves show BBHs with $\xa$ initially
  anti-aligned with $\LN$ ($\theta_1 = 170^\circ, 160^\circ,
  150^\circ$).  The green, blue, and purple curves show BBHs with
  $\xa$ initially aligned with $\LN$ ($\theta_1 = 30^\circ, 20^\circ,
  10^\circ$).}
\label{F:mag3}
\end{figure}

The dotted and solid black curves in the top panel of Fig.~\ref{F:mag1.22} are
identical to within the Poisson noise of our limited number of BBH inspirals,
confirming the finding of
Refs.~\cite{Bogdanovic:2007hp,Herrmann:2009mr,Lousto:2009ka} that isotropic
distributions of BBH spins remain nearly isotropic as they inspiral.
Even at $r_i = 1000 M$, the blue (red) subset of spin configurations yields
the largest (smallest) predicted final spins, because for these configurations
the spin of the more massive black hole is aligned (anti-aligned) with the
orbital angular momentum.  The spin-orbit resonances further enhance (reduce)
the final spins predicted for these subsets by aligning (anti-aligning) the
BBH spins {\it with each other} during the inspiral for small (large) initial
values of $\theta_1$.  As a result, the solid blue (red) distribution at $r_f
= 10 M$ has a larger (smaller) mean final spin than the initial dotted
distribution at $r_i = 1000 M$.  This can be seen in the displacement of
predicted final spins for the colored subsets away from $\chi_f \simeq 0.75$
towards larger and smaller values.

To clarify the magnitude of this effect, we have performed 6
additional sets of BBH inspirals, each of which consists of a fixed
value of $\theta_1$ and a $30 \times 30$ grid evenly spaced in $\cos
\theta_2$ and $\Delta \phi$.  Three of these sets have the spin of the
more massive black hole nearly aligned with the orbital angular
momentum ($\theta_1 = 10^\circ, 20^\circ, 30^\circ$), while the other
3 sets have $\xa$ nearly anti-aligned with $\LN$ ($\theta_1 =
150^\circ, 160^\circ, 170^\circ$).  The choice of aligned
distributions was partly motivated by the finding of
Ref.~\cite{Dotti:2009vz} that accretion torques will align BBH spins
to within $10^\circ$ ($30^\circ$) of the orbital angular momentum for
a cold (hot) disk.  The predicted final spins for these distributions,
both at $r_i = 1000 M$ and $r_f = 10 M$, are shown in the bottom panel
of Fig.~\ref{F:mag1.22}.  The final spins for the initially aligned
($\theta_1 \leq 30^\circ$) BBH distributions are significantly larger
when predicted at $r_f = 10 M$ than at $r_i = 1000 M$, undermining the
claim of \cite{Barausse:2009uz} that the AEIn formula can accurately
predict final spins at large separations without the need for PN
evolutions.  The predicted final spins for the initially anti-aligned
($\theta_1 \geq 150^\circ$) BBH distributions conversely shift to
lower values as the predictions are made later in the inspiral.  We
provide the mean and standard deviation of the final spins predicted
for these 6 new sets of partially aligned BBH distributions for all 5
formulae in Table~\ref{T:fspin}.

To explore the dependence of these effects on the mass ratio, we have provided
histograms of the predicted final spins for these same BBH spin distributions
with $q = 1/3$ in Fig.~\ref{F:mag3}.  The discrete peaks at low values of
$\chi_f$ in the histograms in the top panel are an artifact of the 10 discrete
values of $\cos \theta_1$ in our $10 \times 10 \times 10$ grid.  Each peak
contains 100 points with the same initial value of $\theta_1$.  The decrease
in the width of each peak as the BBHs inspiral from $r_i = 1000 M$ to $r_f =
10 M$ is a consequence of the anti-alignment of the BBH spins for large
$\theta_1$, but the gaps between the peaks would be filled in if we used a
finer grid.  The shifts in the mean values of the peaks should be robust with
respect to the grid spacing.  These shifts for the initially aligned BBH
distributions are provided in Table~\ref{T:fspin} for all 5 formulae for $q =
1/3$, as well as for the intermediate mass ratio $q = 2/3$.

\subsection{Spin Directions} \label{SS:dir}

Before providing quantitative results, we need to clarify what is
meant by the {\it direction} of the spin of the final black hole.  In
what reference frame is this direction defined?  Most of the fitting
formulae calibrated with NR simulations attempt to predict the angle
\be \label{E:thf}
\vartheta_f \equiv \arccos [\hLN(r_f) \cdot \hxf(r_f)]
\ee
between the BBH orbital angular momentum $\LN$ at the separation $r_f
= r_{\rm NR}$ where the NR simulations were performed and the final
spin $\xf$ predicted from the BBH spin configuration at this {\it
same} separation.  The analytical predictions of BKL and Kes were
designed to apply to BBH spin configurations at $r_f = r_{\rm ISCO}$.
If one assumed that neither the orbital angular momentum nor the BBH
spins (upon which the prediction $\hxf(r_f)$ depends) precessed during
the inspiral, one could insert these quantities {\it at any
separation} into the right-hand side of Eq.~(\ref{E:thf}) to predict
$\vartheta_f$.  The angle $\vartheta_f$ is physically interesting
because it quantifies the post-merger alignment between $\xf$ and the
inner edge of the accretion disk if one assumes that torques have
aligned the circumbinary disk with $\LN$.  However, one might also be
interested in the alignment between $\xf$ and a feature like the
galactic disk that is assumed to be aligned with $\LN$ at some larger
scale $r_i$.  In that case, one would need to compute the angle
\be \label{E:thi}
\vartheta_i \equiv \arccos [\hLN(r_i) \cdot \hxf(r_i)]
\ee
between $\LN$ at this larger separation and the final spin $\xf(r_i)$
predicted from the BBH spins at this same separation.

The proper way to predict $\xf$ from the BBH spins at $r_i$ would be to use PN
equations like those specified in Sec.~\ref{S:PN} to propagate those spins and
$\LN$ to down to $r_f$, then insert them into the fitting formula of one's
choice.  The AEIn formula is based on the conjecture that $\xf$ points in the
direction of the total angular momentum $\mathbf{J}$ at {\it any} separation,
since angular momentum is always radiated parallel to $\mathbf{J}$, thus
preserving its direction.  This conjecture is plausible because at large
separations, the precession time $t_p$ is much shorter than the inspiral time
$t_{\rm GW}$.  If the vectors associated with the BBHs precess rapidly enough,
all components except those parallel to $\mathbf{J}$ (which varies on the
longer timescale $t_{\rm GW}$) will average to zero.  The AEIn conjecture is
very useful because it allows $\vartheta_i$ to be computed without solving any
PN equations.  However, the approximation $t_p \ll t_{\rm GW}$ upon which it
depends breaks down at small separations.  This may lead to incomplete
cancellation of the angular momentum radiated perpendicular to $\mathbf{J}$.

\begin{figure}[t!]
\begin{center}
\includegraphics[width=3.5in]{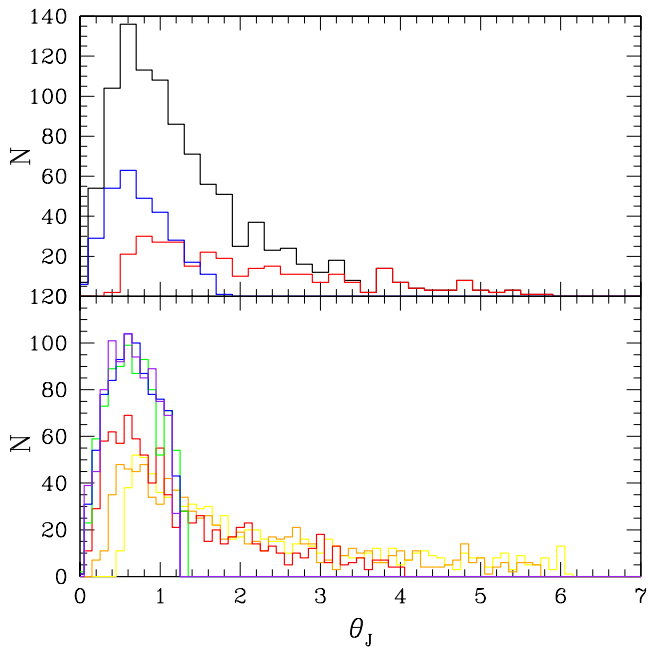}
\end{center}
\caption{{\it Top panel:} Histogram of the angle $\theta_J$ (in
degrees) between the total angular momentum $\mathbf{J}$ at $r_i =
1000 M$ and that at $r_f = 10 M$ for our set of 1000 BBHs with $q =
9/11$ and initially isotropic spins.  As in previous figures, the blue
(red) curve shows the subset of 300 BBHs with the lowest (highest)
initial values of $\theta_1$.  {\it Bottom panel:} Histograms of
$\theta_J$ for the 6 sets of 900 BBH mergers with flat distributions
in $\cos \theta_2$ and $\Delta \phi$ at $r_i = 1000 M$.  The red,
orange, yellow, green, blue, and purple curves show BBHs that have
$\theta_1 = 170^\circ, 160^\circ, 150^\circ, 30^\circ, 20^\circ,$ and
$10^\circ$ respectively at this initial separation.}
\label{F:J1.22}
\end{figure}

\begin{figure}[t!]
\begin{center}
\includegraphics[width=3.5in]{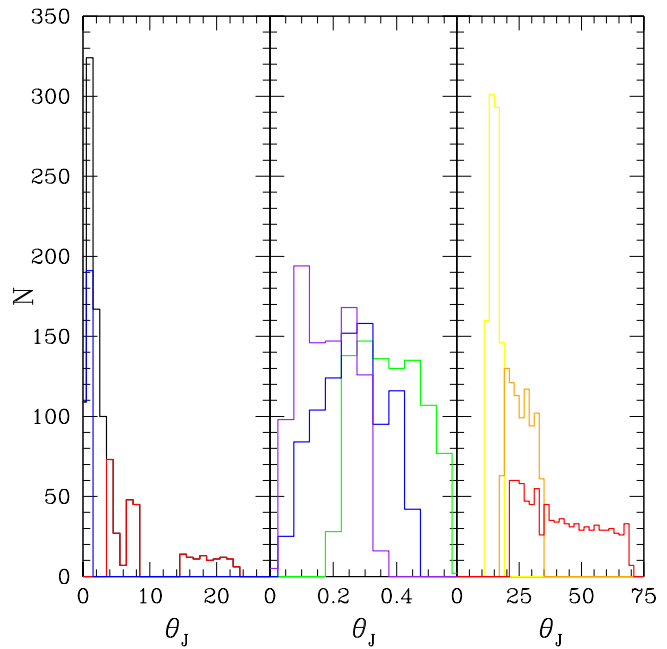}
\end{center}
\caption{{\it Left panel:} Histogram of the angle $\theta_J$ (in
degrees) between the total angular momentum $\mathbf{J}$ at $r_i =
1000 M$ and that at $r_f = 10 M$ for our set of 1000 initially
isotropically spinning BBHs with $q = 1/3$.  As in previous figures,
the blue (red) curve shows the subset of 300 BBHs with the lowest
(highest) initial values of $\theta_1$.  {\it Middle panel:} Histograms
of $\theta_J$ for the 3 sets of 900 BBH mergers initially with
$\theta_1 = 10^\circ$ (purple), $20^\circ$ (blue), and $30^\circ$
(green).  {\it Right panel:} Histograms of $\theta_J$ for the 3 sets
of 900 BBH mergers initially with $\theta_1 = 150^\circ$ (yellow),
$160^\circ$ (orange), and $170^\circ$ (red).}
\label{F:J3}
\end{figure}

We test this possibility by calculating
\be \label{E:thetaJ}
\theta_J \equiv \arccos [\hJ(r_i) \cdot \hJ(r_f)]~,
\ee
the angle between the total angular momentum at $r_i = 1000 M$ and
that after the BBHs have inspiraled to $r_f = 10 M$.  If the direction
of $\mathbf{J}$ really was preserved during the inspiral, $\theta_J$
would vanish.  We present histograms of $\theta_J$ for mass ratio $q =
9/11$ in Fig.~\ref{F:J1.22}.  The upper panel shows the $10 \times 10
\times 10$ grid of BBH spin configurations evenly spaced in ($\cos
\theta_1, \cos \theta_2, \DP$) that we have discussed previously.  The
direction of $\mathbf{J}$ changes by $\theta_J \lesssim 2^\circ$
during most of the inspirals, though a tail extends to larger values
for large initial values of $\theta_1$.  This tail can be seen more
clearly in the bottom panel for the BBHs with $\xa$ initially
anti-aligned with $\LN$ ($\theta_1 \geq 150^\circ$).  We agree with
\cite{Barausse:2009uz} that these large changes in the direction of
$\mathbf{J}$ are likely a consequence of the transitional precession
first identified in Ref.~\cite{Apostolatos:1994mx}.  This transitional
precession occurs to an even greater extent for smaller mass ratios,
as can be seen in Fig.~\ref{F:J3} for $q = 1/3$.  As in the upper
panel of Fig.~\ref{F:mag3}, discrete peaks resulting from the grid
spacing in $\cos \theta_1$ can be seen in the left panel of
Fig.~\ref{F:J3}.  The middle panel shows that the direction of
$\mathbf{J}$ remains nearly constant ($\theta_J \lesssim 0.5^\circ$)
when $\xa$ in closely aligned with $\LN$ ($\theta_1 \leq 30^\circ$).
However, the right panel shows that the assumption of constant $\hJ$
fails badly for the BBHs with $\theta_1 \geq 150^\circ$, that comprise
$\sim 7\%$ of isotropically distributed BBH mergers.  The mass ratio
$q = 1/3$ is not extreme compared to the majority of astrophysical
mergers, so caution should be taken when assuming that $\xf$ points in
the direction of $\mathbf{J}$ such as in Eq.~(\ref{E:thi}).

\begin{figure}[t!]
\begin{center}
\includegraphics[width=3.5in]{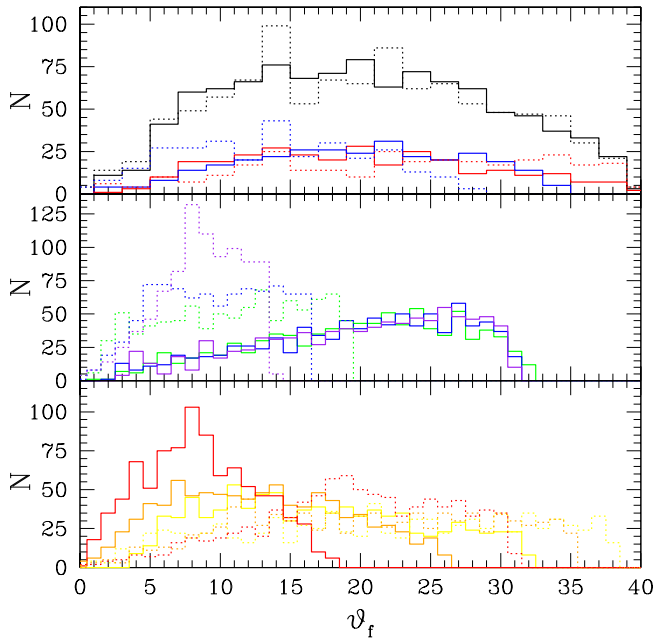}
\end{center}
\caption{{\it Top panel:} Histogram of the angle $\vartheta_f$ (in
degrees) between the orbital angular momentum $\LN$ at $r_f = 10 M$
and the final spin $\xf$ predicted by the AEIn formula from the BBH
spins at that separation.  The BBHs have a mass ratio $q = 9/11$.  As
in previous figures, the black curves show 1000 mergers with initially
isotropic BBH spins, while the blue (red) curves show the subset of
300 BBHs with the lowest (highest) initial values of $\theta_1$.  The
dotted curves show predictions in the absence of spin precession,
while the solid curves show how these predictions change when the BBH
spins precess from $r_i = 1000 M$ to $r_f = 10 M$ according to the PN
equations of Sec.~\ref{S:PN}.  {\it Middle panel:} Histograms of
$\vartheta_f$ for the 3 sets of 900 BBH mergers with $\xa$ initially
aligned with $\LN$ [$\theta_1 = 10^\circ$ (purple), $20^\circ$ (blue),
$30^\circ$ (green)].  {\it Bottom panel:} Histograms of $\vartheta_f$
for the 3 sets of 900 BBH mergers with $\xa$ initially anti-aligned
with $\LN$ [$\theta_1 = 150^\circ$ (yellow), $160^\circ$ (orange),
$170^\circ$ (red)].} \label{F:thf1.22}
\end{figure}

\begin{figure}[t!]
\begin{center}
\includegraphics[width=3.5in]{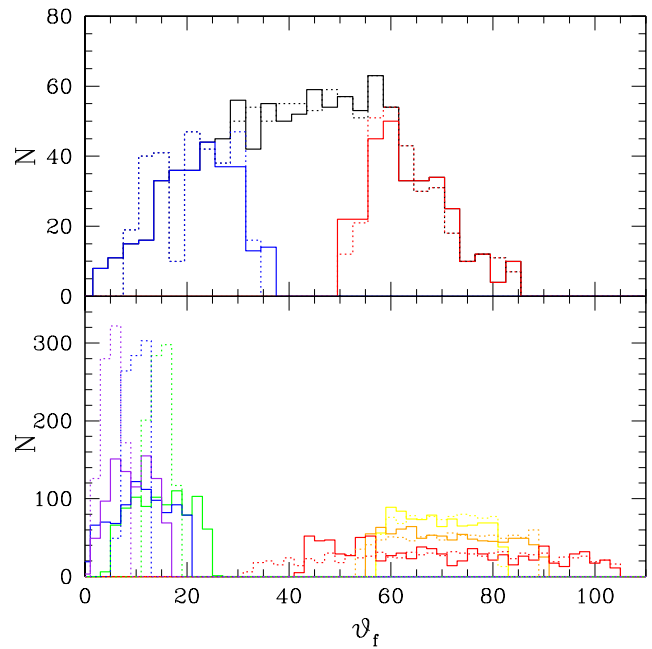}
\end{center}
\caption{Histograms of the angle $\vartheta_f$ predicted by the AEIn
formula for the same sets of BBHs shown in Fig.~\ref{F:thf1.22} but
with a mass ratio $q = 1/3$.  As in that figure, the top panel shows
BBHs with initially isotropic spins with the blue (red) curves
indicating those BBHs with the lowest (highest) initial values of
$\theta_1$.  Dotted curves show predictions without spin precession,
while the solid curves show how these predictions change if the BBHs
spins precess from $r_i = 1000 M$ to $r_f = 10 M$ according to the PN
equations of Sec.~\ref{S:PN}.  The bottom panel shows distributions
with flat initial distributions of $\cos \theta_2$ and $\Delta \phi$,
but with $\theta_1$ now initially set to $170^\circ, 160^\circ,
150^\circ, 30^\circ, 20^\circ,$ and $10^\circ$ respectively for the
red, orange, yellow, green, blue, and purple curves.} \label{F:thf3}
\end{figure}

What about the less ambitious predictions of $\vartheta_f$ from BBH
spins at $r_f = 10 M$, assuming that NR simulations correctly describe
spin precession from this separation until merger?  Spin-orbit
resonances have significant implications for these predictions as
well.  We show predictions of $\vartheta_f$ by the AEIn formula for a
mass ratio of $q = 9/11$ in Fig.~\ref{F:thf1.22}.  The other formulae
predict very similar results.  As in Figs.~\ref{F:mag1.22} and
\ref{F:mag3}, the dotted curves show predictions assuming that the
initial BBH spin distribution is preserved down to $r_f = 10 M$.  The
solid curves include spin precession from $r_i = 1000 M$ to $r_f = 10
M$ according to the PN equations of Sec.~\ref{S:PN}.  The difference
between the dotted and solid black curves in the top panel is below
the Poisson fluctuations, another consequence of the finding of
Refs.~\cite{Bogdanovic:2007hp,Herrmann:2009mr,Lousto:2009ka} that
isotropically oriented BBH spins remain nearly isotropic as they
inspiral.  Careful examination of the upper panel reveals that spin
precession has shifted the BBHs with $\xa$ initially aligned with
$\LN$ (blue distribution) to larger $\vartheta_f$, while the
anti-aligned BBHs have conversely shifted to smaller $\vartheta_f$.

This trend is much more pronounced in the middle and bottom panels of
Fig.~\ref{F:thf1.22}.  Spin precession actually results in the
initially aligned BBHs ($\theta_1 \leq 30^\circ$) having larger values
of $\vartheta_f$ at $r_f = 10 M$ than the anti-aligned BBHs ($\theta_1
\geq 150^\circ$), a reversal of what would be predicted from the
initial spin distributions shown by the dotted curves.  The spin-orbit
resonances explain this highly counterintuitive result.  The BBHs
initially with $\theta_1 \leq 30^\circ$ are influenced by the $\DP =
0^\circ$ resonances which align the BBH spins with each other and
anti-align $\mathbf{S} = \Sa + \Sb$ with $\LN$.  Both effects lead to
larger predicted values of $\vartheta_f$.  Conversely, the BBHs
initially with $\theta_1 \geq 150^\circ$ are influenced by the $\DP =
180^\circ$ resonances, which greatly decrease the magnitude of
$\mathbf{S}$ and align it with $\LN$.  This explains the reduced
values of $\vartheta_f$ for these BBHs seen in the bottom panel of
Fig.~\ref{F:thf1.22}.  This same effect can be seen for a mass ratio
of $q = 1/3$ in Fig.~\ref{F:thf3}, albeit with less significance owing
to the weaker resonances at this smaller mass ratio.
Figs.~\ref{F:thf1.22} and \ref{F:thf3} again illustrate the importance
of accounting for spin precession between $r_i = 1000 M$ and $r_f = 10
M$ when attempting to predict final spins.

\section{Spin Precession Uncertainty} \label{S:err}

So far, we focused on how spin precession between $r_i$ and $r_f$
alters the expected distribution of final spins.  In this Section, we
show that spin precession introduces a fundamental uncertainty in
predicting the final spin.  An uncertainty $\Delta r$ in the BBH
separation leads to an uncertainty $\Delta t_{\rm GW}$ in the time
until merger.  If this uncertainty is comparable to the precession time $t_p$,
the phase of the spin precession at which the merger occurs will be
uncertain as well.  This new uncertainty is independent of and may
exceed that associated with the NR simulations themselves. Readers
only interested in astrophysical distributions of final spins may wish
to proceed to the discussion in Sec.~\ref{S:disc}.

\begin{figure*}[tb]
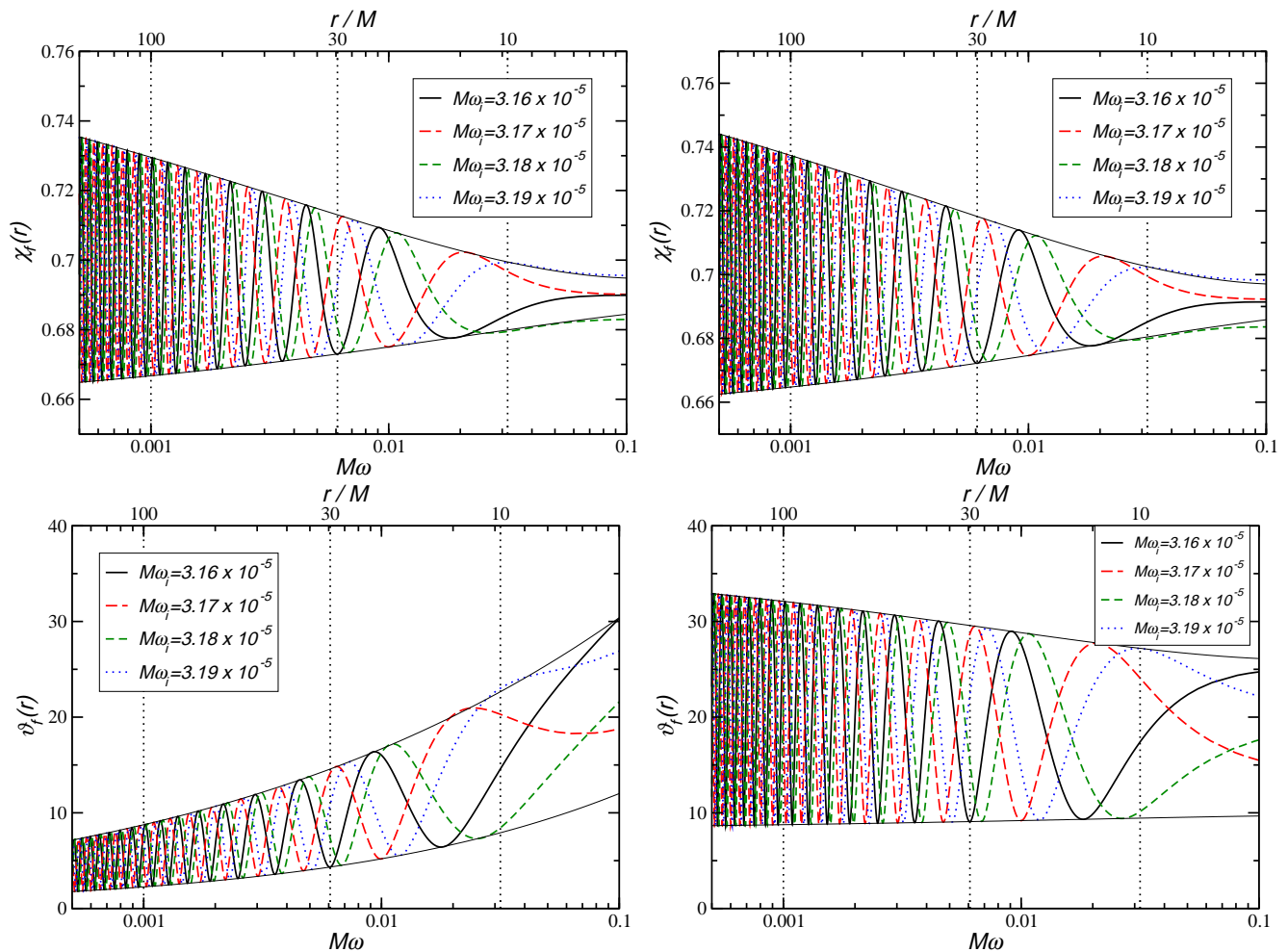

\begin{tabular}{cc}
\includegraphics[scale=0.35,clip=true]{Figure14a.eps}&
\includegraphics[scale=0.35,clip=true]{Figure14b.eps}\\
\includegraphics[scale=0.35,clip=true]{Figure14c.eps}&
\includegraphics[scale=0.35,clip=true]{Figure14d.eps}\\
\end{tabular}
\caption{Predicted $\chi_f$ (upper panels) and $\vartheta_f(r)$ (lower panels)
  obtained from the AEIn (left) and the Kesden (right) formula.  The initial
  parameters of the binary are $q=9/11$, $\chi_1=\chi_2=1$,
  $\theta_1=120^{\circ}$, $\theta_2=60^{\circ}$ and $\Delta \phi=288^{\circ}$.
  The different curves correspond to initial frequencies $M\omega_i=3.16\times
  10^{-5}$ (solid), $3.17\times 10^{-5}$ (long-dashed), $3.18\times 10^{-5}$
  (dashed) and $3.19\times 10^{-5}$ (dotted). The envelope determined for
  $M\omega_i = 3.16\times 10^{-5}$ is displayed by thin solid curves. The
  upper horizontal axis gives the binary separation in units of $M$; the lower
  horizontal axis gives the corresponding orbital frequency $M\omega$.}
\label{fig: illustration_winif}
\end{figure*}

It is often useful to define the final spin direction relative to the
orbital angular momentum $\LN$ at different separations.  We therefore
generalize the angles defined in Eqs.~(\ref{E:thf}) and (\ref{E:thi})
to the separation-dependent quantities
\begin{eqnarray} \label{E:vt}
  \vartheta_f(r) &\equiv& \arccos [\hLN(r) \cdot \hxf(r)]\,, \label{E:vtf} \\
  \vartheta_i(r) &\equiv& \arccos [\hLN(r_i) \cdot \hxf(r)]\,. \label{E:vti}
\end{eqnarray}
Note that these quantities reduce to the previously defined angles in
the appropriate limit: $\vartheta_f(r_f) = \vartheta_f$,
$\vartheta_i(r_i) = \vartheta_i$.  These definitions address two
ambiguities; (i) the choice of the reference orbital angular momentum
and (ii) the separation at which a given fitting formula is evaluated.

\begin{figure*}[tb]
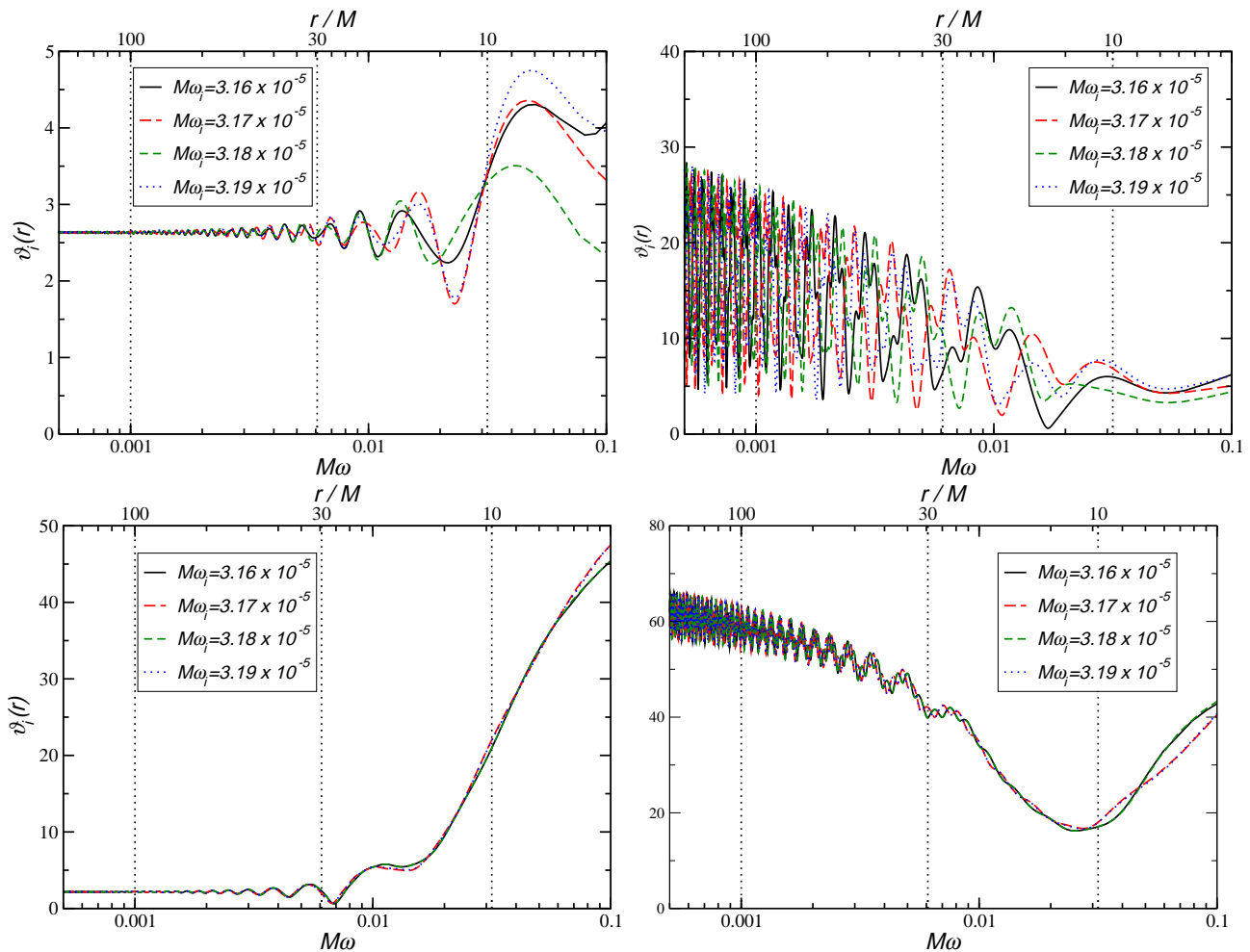

\begin{tabular}{cc}
\includegraphics[scale=0.35,clip=true]{Figure15a.eps}&
\includegraphics[scale=0.35,clip=true]{Figure15b.eps}\\
\includegraphics[scale=0.35,clip=true]{Figure15c.eps}&
\includegraphics[scale=0.35,clip=true]{Figure15d.eps}\\
\end{tabular}
\caption{Predicted spin direction $\vartheta_i(r)$ from the AEIn (left panels)
  and the Kes (right panels) formula.  The upper panel shows the evolution of
  a binary starting with initial parameters $q=9/11$, $\chi_1=\chi_2=1$,
  $\theta_1=120^{\circ}$, $\theta_2=60^{\circ}$ and $\Delta \phi=288^{\circ}$,
  as in Fig.~\ref{fig: illustration_winif}. For comparison, in the bottom
  panels we consider a mass ratio $q=1/3$ and initial spin parameters
  $\chi_1=\chi_2=1$, $\theta_1=154^{\circ}$, $\theta_2=124^{\circ}$ and
  $\Delta \phi=216^{\circ}$. The initial frequency is $M\omega_i=3.16\times
  10^{-5}$ (solid), $3.17\times 10^{-5}$ (long-dashed), $3.18\times 10^{-5}$
  (dashed) and $3.19\times 10^{-5}$ (dotted curve). The upper horizontal axis
  gives the binary separation in units of $M$; the lower horizontal axis gives
  the corresponding orbital frequency $M\omega$.}
\label{fig: illustration_winii}
\end{figure*}

Before we discuss the uncertainties in determining these angles and
the final spin magnitude, we illustrate the evolution of these
quantities during the PN inspiral for a few characteristic examples.
In Fig.~\ref{fig: illustration_winif} we display the final spin
magnitude $\chi_f(r)$ and the angle $\vartheta_f(r)$ as predicted by
the AEIn and the Kesden formulae for a binary with mass ratio
$q=9/11$, extremal spins, and initial spin orientation specified by
the angles $\theta_1=120^{\circ}$, $\theta_2=60^{\circ}$, $\DP =
288^{\circ}$. The behavior of the AEIo, FAU and BKL formulae is quite
similar to the Kesden formula. The different curves in each panel
correspond to slightly different initial frequencies or separations,
$M\omega_i=3.16\times 10^{-5}$, $3.17\times 10^{-5}$, $3.18\times
10^{-5}$ and $3.19\times 10^{-5}$.  The spin precession generically
manifests itself in the oscillatory character of the curves; these
oscillations would be absent for the resonant configurations described
in Sec.~\ref{S:res}.  The thin solid lines represent {\em envelope}
functions obtained by fitting fourth-order polynomials to the maxima
and minima, respectively, of the evolutions starting with
$M\omega_i=3.16\times 10^{-5}$.  Note that these fits contain no
information on the results obtained by using different values of
$M\omega_i$, and yet they still provide excellent envelopes in all cases.

This figure illustrates two ambiguities in predicting $\xf$: (i) the {\it
  initial} frequency $\omega_i$ at which the BBH parameters are specified, and
(ii) the {\it final} separation $r_f$ at which the given formula for $\xf$
should be applied.  Uncertainty in the separation at which the binary
decouples from external interactions could lead to ambiguity in $\omega_i$ in
theoretical studies, while uncertainty in the observed distance, projected
separation, or line-of-sight velocity could lead to uncertainty in $\omega_i$
for models of particular systems.  Gauge-dependent definitions of $r_f$ could
lead to uncertainty in the separation at which fitting formulae should be
applied.  Our task in evaluating the resulting uncertainties for the fitting
formulae AEIn, AEIo, FAU, BKL and Kes introduced in Sec.~\ref{S:dist} is
somewhat simplified because both ambiguities are rooted in the rapid
variations of the phase and in the resulting oscillations in the final
quantities.  These precession-induced oscillations are a clear manifestation
of the hierarchy of time scales introduced in Eq.~(\ref{E:thier}): $t_p \ll
t_{\rm GW}$.

In the upper panels of Fig.~\ref{fig: illustration_winii} we show the angle
$\vartheta_i(r)$ for the same binary configuration illustrated in
Fig.~\ref{fig: illustration_winif}. In the lower panel of Fig.~\ref{fig:
  illustration_winii} we consider instead, for comparison, a system with lower
mass ratio $q=1/3$ and initial spin orientation $\theta_1=154^\circ$,
$\theta_2=124^\circ$, $\Delta \phi=216^\circ$. As before, different curves
correspond to different initial frequencies.  The predicted spin direction as
described by $\vartheta_i(r)$ shows little variation with $\omega_i$.  On the
other hand, the figure demonstrates a strong dependence of $\vartheta_i(r)$ on
the separation $r$ at which we apply the fitting formulae.

\begin{figure*}[htb]
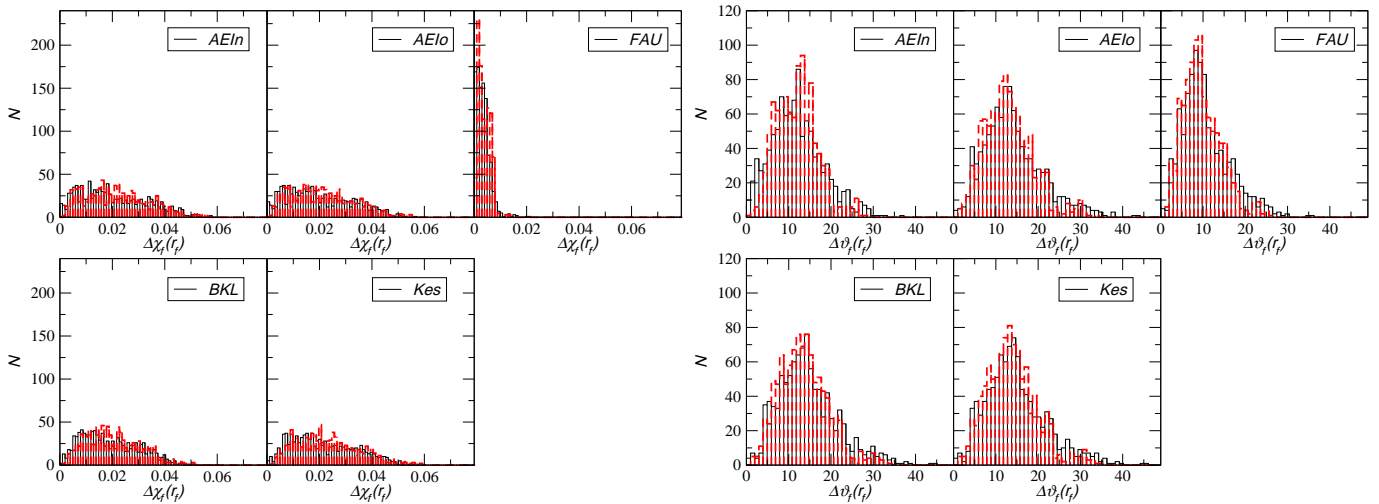

\begin{tabular}{cc}
\includegraphics[scale=0.35,clip=true]{Figure16a.eps}&
\includegraphics[scale=0.35,clip=true]{Figure16b.eps}\\
\end{tabular}
\caption{Uncertainties in the final spin magnitude $\chi_f(r_f)$ (left) and
  direction $\vartheta_f(r_f)$ (right) for extremal BBHs with mass ratio
  $q=2/3$. Solid black histograms were obtained by starting the evolutions at
  $M\omega_i=3.16\times 10^{-5}$ and using the envelope method of
  Fig.~\ref{fig: illustration_winif}. Dashed red histograms were obtained by
  considering the maximum variation in the final quantities as we let
  $M\omega_i$ vary from $3.16\times 10^{-5}$ to $3.22\times 10^{-5}$ in steps
  of $0.005\times 10^{-5}$.  }
\label{fig: q15_wini_vs_env}
\end{figure*}

In the remainder of this Section, we discuss the uncertainties caused
by the rapid spin precession of the following quantities:
\begin{itemize}
  \item [1)] $\chi_f(r_f)$: The magnitude of the final spin as predicted by
    applying a given fitting formula at small binary separation $r_f=10M$,
    i.e. shortly before merger.
  \item [2)] $\vartheta_f(r_f)$: The angle between the orbital angular
    momentum at $r_f=10M$ and the final spin (as predicted using the binary
    parameters at $r_f=10M$).
  \item [3)] $\vartheta_i(r_f)$: The angle between the orbital angular
    momentum of the binary at large separation and the final spin (as
    predicted using the binary parameters at $r_f=10M$).
  \item [4)] $\vartheta_i(r_i)$:  The angle between the orbital angular
    momentum of the binary at large separation and the final spin 
    predicted using the binary parameters at this same large separation.
    We investigate the claim that the AEIn formula, unlike the others, can
    determine this angle  without evolving the BBH parameters down to $r_f$.
\end{itemize}
These quantities are important for modeling the assembly of
supermassive black holes in the context of cosmological structure
formation (see e.g.~\cite{Volonteri:2002vz, Berti:2008af,
Tanaka:2008bv, Lagos:2009xr, Sijacki:2009mn, Fanidakis:2009ct}). They
are also relevant for electromagnetic counterparts of
gravitational-wave sources \cite{Dotti:2006zn}, especially when the
invoked mechanism producing the counterparts depends on the recoil
velocity of the remnant black hole
\cite{Lippai:2008fx,Schnittman:2008ez,Shields:2008va}.

We determine the precession-induced uncertainties as follows. Individual
evolutions, such as those considered in Fig.~\ref{fig: illustration_winif},
suggest that the width of the envelopes or the dispersion induced by varying
the initial frequency provide very similar estimates for the uncertainty in
$\chi_f(r_f)$ and $\vartheta_f(r_f)$.  We have verified this conjecture by
evolving the evenly spaced $10\times 10\times 10$ grid of initially
isotropic, maximally spinning BBH configurations introduced in
Sec.~\ref{S:align} for mass ratio $q=2/3$ and several slightly different
initial frequencies.  When we estimate uncertainties by varying $M\omega_i$
from $3.16\times 10^{-5}$ to $3.22\times 10^{-5}$ in steps of $0.005\times
10^{-5}$ we obtain the red dashed histograms in Fig.~\ref{fig:
  q15_wini_vs_env}. These histograms are in good agreement with the black
solid histograms, where the uncertainty was estimated from the width of the
envelopes. In order to reduce computational cost, in the remainder of this
Section we determine the uncertainties $\Delta \chi_f(r_f)$ and $\Delta
\vartheta_f(r_f)$ by evolving an ensemble of binaries from a {\it single}
initial frequency ($M\omega_i=3.16\times 10^{-5}$) and using the envelope
method.

Fig.~\ref{fig: illustration_winii} shows that the envelope method does
not adequately describe the uncertainty in $\vartheta_i(r)$.  Why does
this angle behave so differently from $\vartheta_f(r)$ as illustrated
in Fig.~\ref{fig: illustration_winif}?  The direction of $\hLN(r_i)$
is fixed, while according to the AEIn formula $\hxf(r)$ points in the
direction of $\hJ(r)$.  The total angular momentum $\mathbf{J}$ only
varies on the radiation timescale $t_{\rm GW}$, so according to
Eq.~(\ref{E:vti}) the AEIn prediction of $\vartheta_i(r)$ should only
vary on this slower timescale as well.  The left panels of
Fig.~\ref{fig: illustration_winii}, at least at small orbital
frequencies $M\omega$ where $t_p \ll t_{\rm GW}$, indeed lack the
high-frequency oscillations characteristic of spin precession.  In
contrast, the Kesden predictions for $\vartheta_i(r)$ shown in the
right panels of Fig.~\ref{fig: illustration_winii} are varying in a
{\it more} complicated way than the predicted values of
$\vartheta_f(r)$.  Changes in the angle $\vartheta_i(r)$ between the
fixed $\hLN(r_i)$ and varying $\hxf(r)$ reflect the full complexity of
spin precession for misaligned, unequal-mass BBHs.  The simpler
variation in $\vartheta_f(r)$ occurs because both $\hLN(r)$ and
$\hxf(r)$ are jointly precessing about $\hJ(r)$, albeit on the same
short timescale $t_p$.

Since the envelope method fails for $\vartheta_i(r)$, we somewhat
arbitrarily define the uncertainty $\Delta \vartheta_i(r_f)$ as the
maximum deviation of $\vartheta_i(r)$ from $\vartheta_i(r_f=10M)$ in
the window $r_f<r<2r_f$. This window covers approximately the range of
initial separations within the reach of present and near-future
numerical relativity simulations, while smaller separations must be
excluded due to the breakdown of the PN expansion.  The formulae other
than AEIn do not claim to predict $\hxf$ from the BBH parameters at
large separations.  To apply these formulae correctly, one must evolve
the BBH parameters inwards to $r_f$ according to PN equations such as
those in Sec.~\ref{S:PN} before applying the formulae.  This evolution
requires significant additional effort, but if performed properly
would only increase $\Delta \vartheta_i(r_i)$ above $\Delta
\vartheta_i(r_f)$ by the uncertainty in the PN equations themselves.
The uncertainty coming from PN evolutions could be quantified by
comparing different PN orders and pushing the calculation of spin
contributions to higher order; such an analysis is beyond the scope of
this paper.  The AEIn formula is special in that it predicts
$\vartheta_i(r_i)$ without this additional PN evolution.  Since AEIn
claims that both $\hLN(r_i)$ and $\hxf(r)$ are independent of $r$, the
uncertainty $\Delta \vartheta_i(r_i)$ for this formula is the maximum
deviation from $\vartheta_i(r_f=10M)$ over the {\it entire} interval
$r_f<r<r_i$.  Since the orbital angular momentum $\LN$ increasingly
dominates over spin contributions in the sum $\mathbf{J} = \LN + \Sa +
\Sb$ at large separations, $\LN$ has little opportunity to precess at
large separations and the uncertainty $\vartheta_i(r_i)$ asymptotes to
a constant value in this limit.

We have evolved the uniform $10\times 10\times 10$ grid of maximally spinning
binaries introduced in Sec.~\ref{S:align} for three different mass ratios:
$q=9/11$, $q=2/3$ and $q=1/3$. The average uncertainties (plus or minus their
associated standard deviations) are summarized in Table
\ref{tab:uncertainties}.

Errors in the final spin magnitudes due to the rapid spin precession
are in the range $\Delta \chi_f \lesssim 0.03$ for all mass
ratios. The FAU formula performs exceptionally well for nearly equal
masses, although it deteriorates to the level of the other predictions
for $q=1/3$.  We suspect that this is because several of the
higher-order terms in $\eta$ in the FAU formula are symmetric in the
dimensionless spins $\xa, \xb$, while physically one would expect the
spin of the more massive black hole to be more important in the limit
$q \to 0$.  Overall however, all formulae are able to predict the spin
magnitude with rather good accuracy.

The uncertainty $\Delta \vartheta_f(r_f)$ in the angle between the
final spin and the orbital angular momentum shortly before merger is
typically in the range of a few to 20 degrees.  Investigation of the
angular dependence of the spin uncertainties shows that the AEIn
formula tends to behave better for initially aligned spins (small
$\theta_1$ and $\theta_2$) and worse for anti-aligned cases.  This is
likely a consequence of anti-aligned binaries being closer to the
limit ${\bf L}(r) \approx -{\bf S}(r)$ where transitional precession
\cite{Apostolatos:1994mx} occurs, violating assumptions (iii) and (iv)
of Ref.~\cite{Barausse:2009uz}.

All formulae are able to predict the angle $\vartheta_i(r_f)$ between
the initial orbital angular momentum and the final spin with decent
accuracy.  The AEIn predictions are overall more accurate, but
investigation of the angular dependence reveals that this accuracy
deteriorates (as expected) when $q=1/3$ and the spin of the larger
black hole is nearly anti-aligned. In this limit the uncertainties
increase up to $\sim 20^\circ$. This is again a consequence of those
configurations approaching the transitional precession regime, where
${\bf L}(r) \approx -{\bf S}(r)$.

The AEIn prediction is unique in that it claims to predict $\vartheta_i(r_i)$
using the binary parameters at large separation without PN evolution. Our
findings confirm (quite remarkably) that the majority of binaries in an
initially isotropic ensemble result in a final spin which is nearly aligned
with the orbital angular momentum at large binary separation.  The values of
$\theta_J$ shown in Figs.~\ref{F:J1.22} and \ref{F:J3} suggest that this would
not be the case for BBHs initially anti-aligned with $\LN$.  The accuracy of
the AEIn predictions also decreases for unequal masses (as expected and
verified by our results for $q=1/3$).  More extreme mass ratios are expected
to play a significant and possibly dominant role in the coalescence of SBH
binaries \cite{Koushiappas:2005qz, Sesana:2007sh, Gergely:2007ny}, so it will
be crucial to test the robustness of the Barausse-Rezzolla predictions for
$q=1/10$ and beyond. Accurate PN evolutions are more difficult in this regime,
and we plan to investigate more extreme mass ratios in the future.

\section{Discussion} \label{S:disc}

In this paper, we examined how precession affects the distribution of
spin orientations as BBHs inspiral from an initial separations $r_i
\approx 1000 M$ where gravitational radiation begins to dominate the
dynamics, all the way down to separations $r_f \simeq 10 M$ where
numerical-relativity simulations typically begin.

We confirmed previous findings that isotropic spin distributions at
$r_i \simeq 1000 M$ remain isotropic at $r_f \simeq 10 M$
\cite{Bogdanovic:2007hp,Herrmann:2009mr,Lousto:2009ka}. However, torques
exerted by circumbinary disks may partially align BBH spins with the
orbital angular momentum at separations $r > r_i$ before gravitational
radiation drives the inspiral \cite{Bogdanovic:2007hp}. Recent
simulations suggest that the residual misalignment of the BBH spins
with their accretion disk could typically be $\sim 10^\circ
(30^\circ)$ for cold (hot) accretion disks, respectively
\cite{Dotti:2009vz}.  Partially motivated by these findings, we carried
out a more careful analysis of spin distributions that are partially
aligned with the orbital angular momentum at $r = r_i$.  We found that
spin precession efficiently aligns the BBH spins with each other when
the spin of the more massive black hole is initially partially aligned
with the orbital angular momentum, increasing the final spin.  We
found the opposite trend when the spin of the more massive black hole
is initially anti-aligned with the orbital angular momentum.  Long
evolutions are necessary to capture the full magnitude of the spin
alignment.  This could explain why these trends were not observed in
the PN evolutions by Lousto {\it et al.} \cite{Lousto:2009ka}, which
began at a fiducial binary separation $r = 50M$.

Some models of BBH evolution (see
e.g.~\cite{Sesana:2007sh,Koushiappas:2005qz}) suggest that SBH mergers
might have comparable mass ratios ($q \lesssim 1$) at high redshift
and more extreme mass ratios at low redshift.  Since spin alignment is
stronger for comparable-mass binaries, more alignment might be
expected in SBH binaries at high redshifts.  Observational arguments
(see e.g.~\cite{Volonteri:2007tu}) and magnetohydrodynamic simulations
of accretion disks \cite{Tchekhovskoy:2009ba} provide some evidence
that black hole spins are related to the radio loudness of quasars.
If so, the inefficient alignment (and consequently smaller spins)
produced by unequal-mass mergers at low redshift would at least be
consistent with recent observational claims that the mean radiative
efficiency of quasars decreases at low redshift
\cite{Wang:2009ws,Li:2010vv}.  Stellar-mass black hole binaries should
also have comparable mass ratios, so significant spin alignment could
occur in such systems as well.

We also pointed out that predictions of the final spin $\xf$ usually
suffer from two sources of uncertainty: (i) the uncertainty in the
{\it initial} frequency $\omega_i$ at which the BBH parameters are
specified, and (ii) the uncertainty in the {\it final} separation
$r_f$ at which the given formula for $\xf$ should be applied. Both
ambiguities are rooted in the rapid precessional modulation of the
orbital parameters, which in turn results from the precessional
timescale $t_p$ being much shorter than the radiation timescale
$t_{\rm GW}$. Spin precession induces an intrinsic inaccuracy $\Delta
\chi_f \lesssim 0.03$ in the dimensionless spin magnitude and
$\Delta \vartheta_f \lesssim 20^\circ$ in the final spin direction.

The spin-orbit resonances studied in this paper should have
significant effects on the distribution of gravitational recoil
velocities resulting from BBH mergers, because the maximum recoil
velocity has a strong dependence on spin alignment
\cite{Gonzalez:2007hi,Campanelli:2007ew,Dotti:2009vz}. We plan to
extend this study to investigate the predictions of different formulae
for the recoil velocities that have been proposed in the literature.

\vspace{0.3cm}

\subsection*{Acknowledgements} 

We are particularly grateful to Vitor Cardoso for helping to test our
numerical implementation of the PN evolution equations described in
Sec.~\ref{S:PN}, and to \'{E}tienne Racine for pointing out the possible relevance
of the quadrupole-monopole interaction.  We would also like to thank Enrico
Barausse, Manuela Campanelli, Yanbei Chen, Pablo Laguna, Carlos Lousto, Samaya
Nissanke, Evan Ochsner, Sterl Phinney and Manuel Tiglio for useful
discussions. This work was supported by grants from the Sherman Fairchild
Foundation to Caltech and by NSF grants No.~PHY-0601459 (PI: Thorne) and
PHY-090003 (TeraGrid).  M.K. acknowledges support from NASA BEFS grant
NNX07AH06G (PI: Phinney). E.B.'s research was supported by NSF grant
PHY-0900735.  U.S. acknowledges support from NSF grant PHY-0652995.

\bibliography{Dec7}

\begin{table*}[h]
  \caption{Mean and standard deviation of the final spin magnitudes predicted
    for different sets of maximally spinning BBH mergers.  The first column
    lists the formulae used to predict the final spins, as described in
    Sec.~\ref{S:dist}.  The second column gives the mass ratio $q$.  Each set
    of BBHs begins at $r_i = 1000 M$ with the indicated value of $\theta_1$
    and flat distributions of $\cos \theta_2$ and $\Delta \phi$.  The third,
    fourth, and fifth columns show the mean and deviation expected if the BBH
    spins do not precess, thus maintaining their initial distributions at $r_i
    = 1000 M$ until merger.  The sixth, seventh, and eighth columns assume
    that the spins precess according to the PN equations of Sec.~\ref{S:PN} as
    they inspiral to $r_f = 10 M$, at which separation we apply the spin
    formulae.}
  \label{T:fspin}
  \begin{ruledtabular}
  \begin{tabular}{l|c|ccc|ccc}
          & & \multicolumn{3}{c|}{$r_i = 1000 M$} & \multicolumn{3}{c}{$r_f = 10 M$} \\
    model & $q$	& $\theta_1=10^\circ$ & $\theta_1=20^\circ$ & $\theta_1=30^\circ$
	  	& $\theta_1=10^\circ$ & $\theta_1=20^\circ$ & $\theta_1=30^\circ$ \\
    \hline
    AEIn & 9/11    	& $0.867\pm 0.064$ & $0.863\pm 0.065$ & $0.857\pm 0.066$
                        & $0.914\pm 0.034$ & $0.905\pm 0.036$ & $0.892\pm 0.038$ \\
    AEIo & 9/11 	& $0.866\pm 0.063$ & $0.863\pm 0.064$ & $0.856\pm 0.065$
                        & $0.912\pm 0.034$ & $0.904\pm 0.036$ & $0.891\pm 0.038$ \\
    FAU  & 9/11 	& $0.873\pm 0.059$ & $0.868\pm 0.060$ & $0.861\pm 0.061$
                        & $0.909\pm 0.035$ & $0.901\pm 0.037$ & $0.888\pm 0.039$ \\
    BKL  & 9/11		& $0.862\pm 0.067$ & $0.858\pm 0.068$ & $0.851\pm 0.070$
                        & $0.905\pm 0.037$ & $0.898\pm 0.039$ & $0.884\pm 0.042$ \\
    Kes  & 9/11		& $0.901\pm 0.072$ & $0.896\pm 0.073$ & $0.889\pm 0.075$
                        & $0.950\pm 0.038$ & $0.941\pm 0.041$ & $0.927\pm 0.044$ \\
    \hline
    AEIn & 2/3    	& $0.886\pm 0.052$ & $0.882\pm 0.053$ & $0.875\pm 0.054$
                        & $0.922\pm 0.030$ & $0.914\pm 0.031$ & $0.900\pm 0.034$ \\
    AEIo & 2/3	 	& $0.886\pm 0.052$ & $0.882\pm 0.052$ & $0.876\pm 0.054$
                        & $0.922\pm 0.030$ & $0.914\pm 0.031$ & $0.900\pm 0.034$ \\
    FAU  & 2/3	 	& $0.901\pm 0.043$ & $0.895\pm 0.044$ & $0.886\pm 0.046$
                        & $0.924\pm 0.029$ & $0.915\pm 0.030$ & $0.901\pm 0.031$ \\
    BKL  & 2/3		& $0.882\pm 0.052$ & $0.878\pm 0.053$ & $0.870\pm 0.054$
                        & $0.914\pm 0.031$ & $0.906\pm 0.032$ & $0.893\pm 0.035$ \\
    Kes  & 2/3		& $0.921\pm 0.056$ & $0.917\pm 0.057$ & $0.909\pm 0.059$
                        & $0.958\pm 0.031$ & $0.949\pm 0.034$ & $0.935\pm 0.037$ \\
    \hline
    AEIn & 1/3    	& $0.950\pm 0.025$ & $0.946\pm 0.025$ & $0.938\pm 0.026$
                        & $0.957\pm 0.023$ & $0.951\pm 0.023$ & $0.941\pm 0.022$ \\
    AEIo & 1/3	 	& $0.958\pm 0.025$ & $0.953\pm 0.026$ & $0.944\pm 0.026$
                        & $0.964\pm 0.023$ & $0.958\pm 0.023$ & $0.947\pm 0.022$ \\
    FAU  & 1/3	 	& $0.972\pm 0.013$ & $0.964\pm 0.014$ & $0.951\pm 0.016$
                        & $0.975\pm 0.012$ & $0.966\pm 0.012$ & $0.953\pm 0.011$ \\
    BKL  & 1/3		& $0.931\pm 0.020$ & $0.927\pm 0.020$ & $0.921\pm 0.021$
                        & $0.936\pm 0.018$ & $0.931\pm 0.018$ & $0.923\pm 0.018$ \\
    Kes  & 1/3		& $0.968\pm 0.021$ & $0.965\pm 0.022$ & $0.958\pm 0.023$
                        & $0.974\pm 0.019$ & $0.970\pm 0.019$ & $0.962\pm 0.020$ \\
    \hline \hline
    model & $q$	& $\theta_1=150^\circ$ & $\theta_1=160^\circ$ & $\theta_1=170^\circ$
	  	& $\theta_1=150^\circ$ & $\theta_1=160^\circ$ & $\theta_1=170^\circ$ \\
    \hline
    AEIn & 9/11    	& $0.551\pm 0.080$ & $0.527\pm 0.080$ & $0.511\pm 0.079$
                        & $0.535\pm 0.072$ & $0.510\pm 0.076$ & $0.493\pm 0.080$ \\
    AEIo & 9/11 	& $0.551\pm 0.080$ & $0.527\pm 0.080$ & $0.512\pm 0.079$
                        & $0.535\pm 0.072$ & $0.510\pm 0.077$ & $0.493\pm 0.080$ \\
    FAU  & 9/11 	& $0.542\pm 0.076$ & $0.520\pm 0.076$ & $0.506\pm 0.076$
                        & $0.530\pm 0.070$ & $0.507\pm 0.074$ & $0.492\pm 0.076$ \\
    BKL  & 9/11		& $0.514\pm 0.088$ & $0.488\pm 0.087$ & $0.471\pm 0.086$
                        & $0.496\pm 0.078$ & $0.468\pm 0.083$ & $0.449\pm 0.087$ \\
    Kes  & 9/11		& $0.531\pm 0.091$ & $0.504\pm 0.090$ & $0.486\pm 0.089$
                        & $0.512\pm 0.081$ & $0.483\pm 0.087$ & $0.463\pm 0.091$ \\
    \hline
    AEIn & 2/3    	& $0.500\pm 0.067$ & $0.467\pm 0.066$ & $0.445\pm 0.065$
                        & $0.490\pm 0.057$ & $0.456\pm 0.062$ & $0.432\pm 0.065$ \\
    AEIo & 2/3	 	& $0.499\pm 0.067$ & $0.466\pm 0.067$ & $0.444\pm 0.066$
                        & $0.489\pm 0.057$ & $0.455\pm 0.062$ & $0.432\pm 0.065$ \\
    FAU  & 2/3	 	& $0.490\pm 0.060$ & $0.460\pm 0.060$ & $0.441\pm 0.059$
                        & $0.483\pm 0.053$ & $0.452\pm 0.056$ & $0.432\pm 0.059$ \\
    BKL  & 2/3		& $0.465\pm 0.072$ & $0.430\pm 0.071$ & $0.405\pm 0.070$
                        & $0.454\pm 0.060$ & $0.416\pm 0.066$ & $0.390\pm 0.070$ \\
    Kes  & 2/3		& $0.480\pm 0.075$ & $0.442\pm 0.071$ & $0.417\pm 0.072$
                        & $0.468\pm 0.063$ & $0.428\pm 0.068$ & $0.401\pm 0.072$ \\
    \hline
    AEIn & 1/3    	& $0.324\pm 0.034$ & $0.233\pm 0.034$ & $0.151\pm 0.032$
                        & $0.323\pm 0.013$ & $0.231\pm 0.015$ & $0.145\pm 0.021$ \\
    AEIo & 1/3	 	& $0.321\pm 0.034$ & $0.230\pm 0.034$ & $0.146\pm 0.032$
                        & $0.319\pm 0.012$ & $0.227\pm 0.015$ & $0.140\pm 0.021$ \\
    FAU  & 1/3	 	& $0.301\pm 0.026$ & $0.222\pm 0.026$ & $0.154\pm 0.024$
                        & $0.300\pm 0.016$ & $0.220\pm 0.018$ & $0.151\pm 0.020$ \\
    BKL  & 1/3		& $0.315\pm 0.034$ & $0.222\pm 0.034$ & $0.136\pm 0.032$
                        & $0.313\pm 0.011$ & $0.219\pm 0.013$ & $0.130\pm 0.019$ \\
    Kes  & 1/3		& $0.322\pm 0.035$ & $0.227\pm 0.035$ & $0.139\pm 0.033$
                        & $0.320\pm 0.012$ & $0.224\pm 0.014$ & $0.133\pm 0.019$ \\
  \end{tabular}
  \end{ruledtabular}
\end{table*}

\begin{table*}[h]
  \caption{Uncertainty distributions in $\chi_f$ and in the various angles
    describing the final spin directions, as predicted by the formulae listed
    in Sec.~\ref{S:dist}. The uncertainties and their standard deviations are
    obtained by evolving uniform $10\times10\times10$ grids of maximally
    spinning BBHs with mass ratio $q=9/11$, $2/3$ and $1/3$, respectively.
    \label{tab:uncertainties}
  }
  \begin{ruledtabular}
  \begin{tabular}{l|c|cc|cc}
    model & $q$ & $\Delta \chi_f(r=10M)$ & $\Delta \vartheta_f(r=10M)$ &
            $\Delta \vartheta_i(r=10M)$ & $\Delta \vartheta_i(r=1000M)$ \\
    \hline
    AEIn & $9/11$ &  $0.0159 \pm 0.0099$  &  $ 8.38 \pm 5.30$  &  $ 1.47 \pm 1.09$  &  $ 1.48 \pm  1.10$  \\
    AEIo & $9/11$ &  $0.0155 \pm 0.0098$  &  $11.38 \pm 6.18$  &  $ 6.55 \pm 2.73$  &  $-$                \\
    FAU  & $9/11$ &  $0.0021 \pm 0.0035$  &  $ 8.51 \pm 4.75$  &  $ 3.67 \pm 1.68$  &  $-$                \\
    BKL  & $9/11$ &  $0.0153 \pm 0.0094$  &  $11.74 \pm 6.39$  &  $ 6.89 \pm 2.92$  &  $-$                \\
    Kes  & $9/11$ &  $0.0174 \pm 0.0105$  &  $11.99 \pm 6.51$  &  $ 7.04 \pm 2.96$  &  $-$                \\
    \hline
    AEIn & $2/3$  &  $0.0205 \pm 0.0127$  &  $11.96 \pm 6.17$  &  $ 1.81 \pm 1.21$  &  $ 1.83 \pm  1.24$  \\
    AEIo & $2/3$  &  $0.0199 \pm 0.0124$  &  $14.10 \pm 6.99$  &  $ 7.37 \pm 2.80$  &  $-$                \\
    FAU  & $2/3$  &  $0.0034 \pm 0.0026$  &  $10.66 \pm 5.61$  &  $ 4.41 \pm 1.76$  &  $-$                \\
    BKL  & $2/3$  &  $0.0191 \pm 0.0108$  &  $14.52 \pm 7.05$  &  $ 7.79 \pm 2.98$  &  $-$                \\
    Kes  & $2/3$  &  $0.0217 \pm 0.0124$  &  $14.83 \pm 7.24$  &  $ 8.02 \pm 2.99$  &  $-$                \\
    \hline
    AEIn & $1/3$  &  $0.0165 \pm 0.0109$  &  $ 8.58 \pm 4.17$  &  $ 3.96 \pm 4.46$  &  $ 4.25 \pm  5.16$  \\
    AEIo & $1/3$  &  $0.0156 \pm 0.0101$  &  $ 9.57 \pm 4.62$  &  $10.45 \pm 4.12$  &  $-$                \\
    FAU  & $1/3$  &  $0.0177 \pm 0.0090$  &  $ 7.80 \pm 3.84$  &  $ 6.60 \pm 2.75$  &  $-$                \\
    BKL  & $1/3$  &  $0.0148 \pm 0.0089$  &  $ 9.81 \pm 4.79$  &  $11.24 \pm 4.45$  &  $-$                \\
    Kes  & $1/3$  &  $0.0167 \pm 0.0105$  &  $10.01 \pm 5.06$  &  $11.49 \pm 4.48$  &  $-$                \\
  \end{tabular}
  \end{ruledtabular}
\end{table*}

\end{document}